\documentclass[%
 superscriptaddress,
 reprint,
 showpacs,preprintnumbers,
 nofootinbib,
 amsmath,amssymb,
 aps,
 prd,
 floatfix,
]{revtex4-1}

\usepackage{graphicx}% Include figure files

\usepackage{xcolor}
\usepackage{lineno}
\usepackage{float}
\usepackage{afterpage}
\usepackage{lipsum}

\usepackage{amssymb,amsmath,amstext,amsthm,amsfonts,amsfonts}
\usepackage[utf8]{inputenc}
\usepackage{url}
\usepackage{soul}
\usepackage{bm}
\usepackage[normalem]{ulem}
\usepackage{multirow, makecell}
\usepackage{booktabs}

\usepackage{algorithm}
\usepackage{algpseudocode}
\algrenewcommand\algorithmicrequire{\textbf{Input:}}
\algrenewcommand\algorithmicensure{\textbf{Output:}}
\algnewcommand{\LineComment}[1]{\State \(\triangleright\) #1}
\algrenewcommand{\algorithmiccomment}[1]{\hfill \(\triangleright\) \texttt{\small #1}}

\usepackage{xspace}
\newcommand{\lucid}{{LUCiD}\xspace}
\makeatletter
\newcommand{\algcaption}[1]{%
  \refstepcounter{algorithm}%
  \hrule\vspace{0.6ex}
  \noindent\textbf{Algorithm~\thealgorithm.}\ #1\par
  \vspace{0.6ex}\hrule\vspace{1ex}
}
\makeatother

\usepackage{hyperref}
\hypersetup{
    hidelinks,               % Removes all borders/boxes
    colorlinks=true,         % Enables colored links
    citecolor=teal,          % Color for citations
    linkcolor=teal,           % Color for internal links
    urlcolor=teal           % External URLs
}

\begin{document}

\author{Omar Alterkait}
\email{Omar.Alterkait@tufts.edu}
\affiliation{Department of Physics and Astronomy, Tufts University, 574 Boston Ave, Medford, MA 02155, USA}
\affiliation{The NSF AI Institute for Fundamental Interactions, MIT, Cambridge, MA 02139, USA}

\author{C\'esar Jes\'us-Valls}
\email{Cesar.Jesus@cern.ch}
\affiliation{European Organization for Nuclear Research (CERN), 1211 Geneva 23, Switzerland}

\author{Ryo Matsumoto}
\affiliation{Department of Physics, Institute of Science Tokyo, Meguro, Tokyo 152-8551, Japan}

\author{Patrick de Perio}
\affiliation{Center for Data-Driven Discovery, Kavli IPMU (WPI), UTIAS, The University of Tokyo, Kashiwa, Chiba 277-8583, Japan}

\author{Kazuhiro Terao}
\affiliation{SLAC National Accelerator Laboratory, Menlo Park, CA, 94025, USA}

\title{End-to-end Differentiable Calibration and Reconstruction for Optical Particle Detectors}

\begin{abstract}
\noindent Large-scale homogeneous detectors with optical readouts are widely used in particle detection, with Cherenkov and scintillator neutrino detectors as prominent examples. Analyses in experimental physics rely on high-fidelity simulators to translate sensor-level information into physical quantities of interest. This task critically depends on accurate calibration, which aligns simulation behavior with real detector data, and on tracking, which infers particle properties from optical signals. We present the first end-to-end differentiable optical particle detector simulator, enabling simultaneous calibration and reconstruction through gradient-based optimization. Our approach unifies simulation, calibration, and tracking, which are traditionally treated as separate problems, within a single differentiable framework. We demonstrate that it achieves smooth and physically meaningful gradients across all key stages of light generation, propagation, and detection while maintaining computational efficiency. We show that gradient-based calibration and reconstruction greatly simplify existing analysis pipelines while matching or surpassing the performance of conventional non-differentiable methods in both accuracy and speed. Moreover, the framework's modularity allows straightforward adaptation to diverse detector geometries and target materials, providing a flexible foundation for experiment design and optimization. The results demonstrate the readiness of this technique for adoption in current and future optical detector experiments, establishing a new paradigm for simulation and reconstruction in particle physics.

\end{abstract}
\maketitle

\section{Introduction}
Large-scale homogeneous particle detectors with optical readouts are a cornerstone of modern particle physics research. Prominent examples include neutrino detectors such as Super-Kamiokande~\cite{Super-Kamiokande:2002weg}, SNO~\cite{SNO:1999crp}, KamLAND~\cite{Piepke:2001tg}, Borexino~\cite{Borexino:2008gab}, Daya Bay~\cite{DayaBay:2015kir}, IceCube~\cite{IceCube:2016zyt} and KM3Net~\cite{KM3Net:2016zxf} and JUNO~\cite{JUNO:2015zny}. This technology has critically shaped our modern understanding of neutrino physics~\cite{Kamiokande-II:1987idp, Super-Kamiokande:1998kpq,Super-Kamiokande:2001ljr,SNO:2002tuh,SNO:2001kpb,KamLAND:2002uet,KamLAND:2004mhv,DayaBay:2012fng,BOREXINO:2014pcl,BOREXINO:2020aww,Borexino:2010dli,T2K:2011ypd,T2K:2019bcf, IceCube:2013low, IceCube:2018cha, KM3NeT:2025npi, JUNO:2025gmd, Esteban:2024eli}. Several of these experiments remain operational or are entering new phases with enhanced capabilities, and are expected to continue delivering leading results in the coming decades. In parallel, next-generation experiments such as Hyper-Kamiokande~\cite{Hyper-Kamiokande:2018ofw} and proposed technologies for DUNE Phase~II~\cite{Fiza:2025zfo} will further extend the reach of this detector paradigm, collectively shaping the future of the field.\\

Like most other particle detectors, these types of detectors rely on physics simulators to translate detector-level information, such as sensor readings, into physics observables, such as the type of observed particles and their kinematics. The conventional procedure consists of three strongly co-dependent parts:
\begin{itemize}\setlength\itemsep{-0.3em}
\item \textbf{Simulation: } A digital replica of the experiment aiming to generate detector-level information analogous to the real detector.
\item \textbf{Calibration: } A method that utilizes data collected under well-known conditions to align the simulation description to the detector behavior.
\item \textbf{Reconstruction: } An algorithm that utilizes simulation predictions to infer physics observables from detector-level data.
\end{itemize}
A persistent challenge in particle physics is that existing detector simulators are designed as prediction-only tools without built-in parameter optimization, necessary for calibration and reconstruction tasks. The prevalence of this approach is illustrated by the fact that GEANT4~\cite{GEANT4:2002zbu}, the simulation toolkit used to develop such simulators, is the most cited\footnote{According to the INSPIRE HEP portal.} article in experimental particle physics. The state-of-the-art methodology is, and has been for decades, to develop separate calibration and reconstruction solutions utilizing the detector simulation as needed, typically in a sequence of disjoint steps. This approach introduces artificial task divisions that contribute to inefficient workflows and prevent fully accounting for parameter correlations that contribute to data-simulation mismatches and poor understanding of the underlying physics processes. This affects the ultimate experimental measurement robustness and sensitivity.

In this work, we demonstrate how a differentiable simulator bridges the gap between detector modeling and data analysis, transforming traditionally discrete and independent procedures into a single, continuous, and optimizable pipeline. Focusing on optical detectors, we showcase the first end-to-end differentiable simulator combining the necessary physics fidelity and computational efficiency to be readily applicable to existing and planned physics experiments.

This work represents a significant step toward the establishment of a new paradigm in experimental particle physics, with the potential to supersede existing approaches in terms of simplicity, efficiency, and robustness, while enabling new strategies for detector design optimization and uncertainty estimation.

\subsection{Differentiable Simulations}

Data analysis in particle physics is fundamentally an inverse problem. Given a set of observed sensor readouts, one seeks to infer the latent physical parameters, such as particle momenta or detector properties, that generated them. In the standard paradigm, this inference relies on forward simulations like \textsc{Geant4}~\cite{GEANT4:2002zbu} to maximize a likelihood or minimize a loss function. Since standard simulators function as non-differentiable black boxes, this optimization typically relies on discrete grid searches, derivative-free algorithms, or finite-difference approximations.

This approach suffers severely from the \textit{curse of dimensionality}. As the number of parameters increases, the volume of the search space grows exponentially. Computing gradients via finite differences becomes prohibitively expensive because it requires running the full simulation at least once for every single parameter. Consequently, experimental workflows are forced to adopt a sequential calibration strategy. Subsets of parameters, such as geometry, optical properties, and electronics response, are tuned independently in a fixed order. This sequential treatment ignores the complex correlations between parameters. For instance, a bias in the absorption length might be compensated by a shift in photosensor efficiency, leading to suboptimal working points and systematic uncertainties that are difficult to quantify.

Differentiable programming offers a robust solution to these challenges by rendering the simulation code itself differentiable. This paradigm leverages Automatic Differentiation (AD)~\cite{Baydin_2018}, a technique distinct from both symbolic differentiation and numerical finite differences. AD decomposes a computer program into a sequence of elementary operations and applies the chain rule of calculus to propagate and combine partial derivatives through each operation in order to obtain the overall derivative. This allows the exact gradient of the simulation output with respect to any input parameter to be computed efficiently. Crucially, in reverse-mode AD, the computational cost of evaluating the full gradient vector is roughly constant regardless of the number of input parameters.

A powerful feature of this paradigm is the ability to construct hybrid models. Well-understood physical laws, such as ray optics, can be implemented as differentiable operators to maintain interpretability, while complex or computationally expensive microscopic processes can be approximated by neural networks. This allows the simulation to retain the expressiveness and efficiency of machine learning without sacrificing the physics processes of the experiment.

The adoption of differentiable programming is rapidly expanding within high-energy physics. The MODE collaboration has formalized the potential of this paradigm for end-to-end optimization, proposing that hardware design and reconstruction algorithms should be optimized simultaneously to maximize physics sensitivity~\cite{Dorigo_2023}. At the fundamental physics level, tools like \textsc{MadJax} have introduced differentiable matrix elements to incorporate theory parameters directly into gradient-based pipelines~\cite{Heinrich_2023}. Similar techniques have been applied to jet physics to enable gradient-based variational inference over latent clustering histories~\cite{Macaluso_2021}, and to statistical analysis pipelines to optimize selection cuts directly for significance~\cite{Simpson_2023}.

In the domain of detector simulation, significant progress has been made for Liquid Argon Time Projection Chambers (LArTPCs). Recent work has demonstrated a fully differentiable LArTPC simulator capable of simultaneously calibrating entangled detector parameters, such as electron drift and signal induction, using gradient descent~\cite{Gasiorowski_2024}.

This work is the first to focus on large-scale homogeneous optical detectors. To the best of our knowledge, it presents the first end-to-end differentiable simulation framework capable of jointly addressing calibration and reconstruction tasks within a single optimization pipeline, while achieving competitive computational efficiency and physics performance relative to established non-differentiable approaches.

\subsection{Optical Particle Detectors}

Most optical particle detectors rely on the detection and characterization of either Cherenkov light, scintillation light, or a combination of both. Cherenkov radiation occurs when charged particles traverse a medium with velocity $v$ exceeding the local phase velocity of light, defined by $v > c/n$, where $n$ is the refractive index. This process produces photons emitted instantaneously along a cone with half-opening angle $\theta_c$ satisfying $\cos\theta_c = c/(nv)$. Scintillation light, in contrast, is emitted isotropically, with an intensity and time profile that depends on the local energy deposition and the detector medium properties.

Optical detectors measure the spatial and temporal distribution of emitted photons using large volumes of transparent material instrumented with arrays of photosensors. This technology has proven particularly powerful in neutrino physics, enabling the construction of large, homogeneous, and cost-effective detectors.

Cherenkov detectors thereby identify the final-state particle species above threshold and reconstruct their kinematics, in particular the flavor and momentum of the charged leptons produced in neutrino interactions, whereas scintillation detectors act primarily as calorimeters with precise timing, sensitive to all particles depositing energy but largely blind to their individual trajectories. Together, these capabilities have been central to the study of neutrinos from accelerator, reactor, atmospheric, solar, supernova, and high-energy astrophysical sources.

\section{End-to-End Differentiable Simulation}

We developed a JAX-based open-source \textbf{L}ight-based \textbf{U}nified \textbf{C}alibration and track\textbf{i}ng \textbf{D}ifferentiable simulation, \textbf{\lucid}\footnote{\href{https://github.com/CIDeR-ML/LUCiD/}{https://github.com/CIDeR-ML/LUCiD/}}.

LUCiD serves as a demonstrator of an end-to-end differentiable simulation. It provides utilities to describe the detector geometry, photon generation, propagation through the detector medium, and hit formation while preserving differentiability throughout the entire pipeline. This section reviews these core components. In Sections~\ref{sec:calibration} and~\ref{sec:tracking}, we then illustrate how gradients drive calibration and track reconstruction. Both are solved with the same second-order Fisher--Gauss--Newton optimizer, described in Sec.~\ref{sec:gauss_newton}.

JAX was chosen as the core framework guided by the intention to build a robust and flexible system capable of accommodating the varied needs of different experiments and serving as a common tool for the broader community. Its automatic differentiation capabilities enable the exact computation of gradients through arbitrarily complex operations, a crucial feature for end-to-end optimization. JAX also provides just-in-time compilation and vectorization, ensuring efficient execution on both CPUs and GPUs while supporting large-scale parallelization. Its functional programming paradigm and composable design make it straightforward to adapt and extend to the specific needs of individual experiments.

\subsection{Detector Geometry}

% \begin{figure}[htbp]
% \centering
%     \caption{Examples of three detector geometries built parametrically using LUCiD.}
%     \includegraphics[width=0.15\textwidth,trim={400pt 350pt 400pt 500pt},clip]{Figures/cylinder_example.pdf}
%     \hfill
%     \includegraphics[width=0.15\textwidth,trim={650pt 650pt 650pt 650pt},clip]{Figures/sphere_example.pdf}
%     \hfill
%     \includegraphics[width=0.15\textwidth,trim={100pt 00pt 0pt 600pt},clip]{Figures/box_example.pdf}
%     \label{fig:det_geom}

% \end{figure}

\begin{figure}[htbp]
\centering
    \caption{Examples of four detector geometries built parametrically using LUCiD.}
    \includegraphics[width=0.48\textwidth]{Figures/geometry_gallery.pdf}
    \label{fig:det_geom}
\end{figure}

We consider optical detectors defined by a physical volume instrumented with photosensors in arbitrary 3D arrangements. We model the four most common geometries used in current and future optical neutrino detectors: cylinders, spheres, and boxes with photosensors on the detector surface, as well as large-scale volumetric string-based detectors, all with arbitrary size, material, and proportions.

The principles that ensure differentiability throughout the entire simulation chain are inherently geometry-agnostic. Therefore, extending this framework to other detector geometries generally does not require any conceptual modification, but rather bespoke configurations of the existing detector templates. When needed, the modular structure of \lucid also allows these templates to be extended to match individual use-cases.

This generality provides a solid foundation for developing a community-wide tool that can be readily adopted across experiments. Moreover, the differentiable nature of the simulation inherently enables built-in reconstruction for any detector configuration, naturally supporting efficient detector design optimization, as discussed in Sec.~\ref{sec:det_design}.

Figure~\ref{fig:det_geom} illustrates four geometry types implemented in \lucid. Unless otherwise specified, results are presented for a cylindrical detector with parameters $R = 16.9$~m, $H = 36.2$~m, $s_R =25$~cm, $N_s = 10764$, and $c_{\mathrm{med}}$, corresponding respectively to the cylinder radius, height, sensor radius, number of sensors, and the speed of light in the detector medium. In this work, the medium is water, and we take $c_{\mathrm{med}} = c / n$ with a refractive index $n = 1.33$. These values approximate the geometry of the Super-Kamiokande~\cite{Super-Kamiokande:2002weg} detector; accordingly, we refer to this configuration as \textit{SK-like}.

\subsection{Modeling Expected Detector Responses}
For calibration and reconstruction, \lucid is designed to produce smoothed, differentiable expectations in detector-output space: predicted sensor responses that vary continuously with the underlying physical parameters and are suitable for gradient-based inference.

It is important to emphasize a conceptual distinction in objectives. Monte Carlo toolkits such as \textsc{Geant4} aim to generate individual events whose statistical properties match those of measured data by sampling physical photons and microscopic processes. By contrast, our simulator's target quantity is the \emph{expected} detector response (an ``average event'') rather than individual stochastic realizations.

To represent these expectations we use \emph{rays}, mathematical objects that encode averaged contributions (or superpositions) of many possible photon trajectories and outcomes. While it would be possible, in principle, to retain differentiability with an explicit photon-sampling approach, we choose rays because they directly represent the ensemble expectation we want to predict and because they facilitate variance-reduction during propagation, later discussed in Sec.~\ref{sec:photon_prop}. Each ray therefore need not correspond to a single physical photon; instead it carries an intensity and parameterization that together approximate the expected contribution of many photons. The intensity represents a fractional photon count. The user-specified parameter \(N_{\rm rays}\) controls the granularity of this approximation, and individual ray intensities are automatically scaled to preserve the overall normalization of the expected signal. In this parameterization, the simulation approaches the continuous wave propagation limit as the number of rays approaches infinity.

This formulation does not alter the underlying physics that determines the optimal parameter values. In the high-statistics limit, a conventional sampling-based simulation would yield the same expected detector response and, consequently, the same loss minimum. The techniques introduced in subsequent sections serve distinct roles in making gradient-based optimization practical: implicit capture (Sec.\ref{sec:photon_prop}) provides variance reduction, allowing the expectation to be estimated accurately with substantially fewer rays, while photon relaxation (Sec.\ref{sec:photon_relaxation}) addresses discontinuities inherent to geometric ray-sensor intersection, ensuring that gradients remain well-defined and physically meaningful. The fundamental advantage of the differentiable framework is therefore twofold: reduced computational cost to evaluate the loss surface, and access to exact gradients that enable efficient optimization in high-dimensional parameter spaces.

\subsection{Ray Generation}
\label{sec:ray_generation}

The goal of the ray-generation step is to construct a differentiable function that maps a set of physical parameters describing a charged-particle track or calibration configuration to a list of rays characterized by their emission position $\mathbf{x}_\gamma$, direction $\mathbf{u}_\gamma$, wavelength $\lambda_\gamma$, and time $t_\gamma$. Formally, we define
\begin{equation}
f_{\gamma}(\mathbf{\theta}_{\mathrm{event}}, \mathbf{\theta}_{\mathrm{detector}})
\rightarrow \{\mathbf{x}_i, \mathbf{u}_i, \lambda_i, t_i\}_{i=1}^{N_\gamma},
\end{equation}
where $\mathbf{\theta}_{\mathrm{event}}$ and $\mathbf{\theta}_{\mathrm{detector}}$ denote the event and detector parameters, respectively.

One approach to define $f_{\gamma}$ is to describe all effects analytically using differentiable primitives, i.e. continuous mathematical operations that replicate the underlying physics while preserving gradient flow, an avenue we follow later when modeling ray propagation (see Sec.~\ref{sec:photon_prop}). However, re-implementing the complete set of microscopic effects is unnecessary for our use case. Since the physics governing the passage of particles through matter is well established and does not need to be optimized, representing $f_{\gamma}$ through a surrogate model simplifies the implementation complexity and provides an opportunity to illustrate how differentiable simulators can be built combining analytical parametrizations and surrogate models.

We construct a differentiable surrogate for the Cherenkov light emission $f^{\mathrm{Cherenkov}}_\gamma(E, \theta, s)$, which reproduces the expected photon emission distributions as a function of the initial particle energy $E$, photon emission distance along the track $s$, and emission angle $\theta$. This parametrization is analogous to that used in FitQun~\cite{Patterson:2009ki, Super-Kamiokande:2019gzr}, an official reconstruction software of the T2K, Super-Kamiokande and Hyper-Kamiokande experiments. To tune our surrogate, we first generate 3D lookup tables of $(E, \cos\theta, \tilde{s})$ distributions using \textsc{PhotonSim}\footnote{\href{https://github.com/cesarjesusvalls/PhotonSim/}{https://github.com/cesarjesusvalls/PhotonSim/}}, a GEANT4-based utility developed for this purpose, and then train a Sinusoidal Representation Network (SIREN)~\cite{sitzmann2020implicit} to interpolate them. We use the normalized distance $\tilde{s} \equiv s/s_{\max}(E)$, where $s_{\max}(E)$ is an energy-dependent characteristic track length: a high quantile ($99.99\%$) of the longitudinal photon-emission-distance distribution, inflated by a small safety factor so that essentially all of the emission falls within $\tilde{s}\in[0,1]$. Its energy dependence is fit per particle species with a smooth, monotonic form: a power law $s_{\max}(E)=A\,E^{B}$ for muons, and a piecewise power law for electrons, needed to describe $s_{\max}$ at low energies, so that a single surrogate spans a wide range of particle energies, from energies near the Cherenkov threshold up to $100$~GeV. Details of the model architecture and training procedure are provided in Appendix~\ref{app:siren_training}.

% For each ray $i$, random values of $\theta_i$ and $s_i$ are drawn from uniform intervals\footnote{Since $f^{\mathrm{Cherenkov}}$ is quite sparse, we utilize an adaptive binning strategy to avoid sampling in regions where the intensities are zero, further details are presented in Appendix~\ref{app:adaptive_binning}.}, and a corresponding generation intensity
% \begin{equation}
% I^{\mathrm{generation}}_i = f^{\mathrm{Cherenkov}}_\gamma(E, \theta_i, s_i)
% \end{equation}
% is assigned. This intensity is carried throughout the simulation chain, ensuring that the detector response remains differentiable with respect to the input particle energy.

For each ray $i$, the emission angle $\theta_i$ and normalized distance $\tilde{s}_i$ are obtained by two-pass importance sampling of the surrogate. In the first pass the SIREN is evaluated on a regular grid over the $(\theta,\tilde{s})$ domain, yielding a discrete proposal density $q$. Each ray then selects a grid cell with probability proportional to $q$, and is placed within that cell by a Latin-hypercube-stratified jitter, so that rays sharing a cell cover its area uniformly. In the second pass the SIREN is evaluated at the jittered points, and each ray is assigned the importance weight
\begin{equation}
w_i = \frac{f^{\mathrm{Cherenkov}}_\gamma(E, \theta_i, \tilde{s}_i)}{q_i},
\end{equation}
which divides out the proposal and keeps the per-ray weights of order unity across the whole domain. Because the Cherenkov emission is sharply peaked, drawing the cell from the surrogate density concentrates the rays where the emission is bright while retaining the full phase space, including the low-intensity wide-angle tail.

The generation intensity follows as the normalized weight scaled to the absolute photon yield,
\begin{equation}
I^{\mathrm{generation}}_i = \frac{w_i}{\sum_j w_j}\, N_\gamma(E),
\end{equation}
so that $\sum_i I^{\mathrm{generation}}_i = N_\gamma(E)$ by construction: the SIREN supplies the shape of the emission and the stored $N_\gamma(E)$ power law its absolute scale. The normalized distance is converted to a physical distance along the track as $s_i = \tilde{s}_i\,s_{\max}(E)$. This intensity is carried throughout the simulation chain, ensuring that the detector response remains differentiable with respect to the input particle energy.

With $\theta_i$ and $s_i$ known for every ray, we compute the ray origin and direction as a function of the track parameters $\mathbf{x}_{\rm track}$ and $\mathbf{u}_{\rm track}$:
\begin{equation}
    \mathbf{x}_\gamma = \mathbf{x}_{\rm track} + s_i\,\mathbf{u}_{\rm track},
\end{equation}
\begin{equation}
    \mathbf{u}_\gamma = \cos\theta_i\,\mathbf{u}_{\rm track}
    + \sin\theta_i\left( \cos\phi_i\,\hat{v} + \sin\phi_i\,\hat{w} \right),
\end{equation}
where $\phi_i$ is a random azimuthal angle uniformly distributed in $[0, 2\pi)$, and $\hat{v}$ and $\hat{w}$ are orthonormal vectors perpendicular to $\mathbf{u}_{\rm track}$ defining the local emission plane. These expressions correspond to rays emitted along a cone of opening angle $\theta_i$ around the particle trajectory, at a distance $s_i$ from the track origin. Through these expressions, the rays predictions become continuous functions ensuring that the detector response remains differentiable with respect to the track origin and direction.

This approach requires one surrogate model per particle type and material of interest, resulting in a small set of lookup tables that need to be generated only once\footnote{\textsc{PhotonSim} and \lucid provide open-source utilities to generate the required GEANT4 photon samples, fit the surrogate models, and construct the corresponding emission-time parameterizations described later.}. In this work we construct surrogates for muons and electrons. Each surrogate includes the average light of the associated secondary cascade with the exception of the delayed Michel electron from muon decay, which is well separated in time and can be treated in isolation. Surrogates for other particles, such as pions and protons, could be constructed analogously by introducing additional logic to handle their inelastic interactions (see the discussion in Secs.~\ref{sec:future_dir_topos} and~\ref{sec:future_dir_stochasticity}). In practice, however, lepton-only surrogates are sufficient for multi-particle reconstruction: the lepton is identified as the single-particle pattern that best matches the observed light, providing trustworthy kinematic estimates, while any remaining activity that departs from the lepton expectation is labeled as hadronic. This categorization is sufficient to reproduce the single- and multi-ring samples used, under their usual lepton-based kinematic binning, in the T2K and Super-Kamiokande oscillation analyses.

The initial emission time $t_0$ can be described by the analytical parameterization

\begin{equation}
\label{eq:t0_correction}
f^{t_0}_\gamma(E,s) = A(E)\left[\exp\!\left(\left(\frac{s}{\lambda(E)}\right)^{\beta(E)}\right) - 1\right],
\end{equation}

a stretched exponential that passes through the origin and whose energy-dependent coefficients are fitted to predictions from \textsc{PhotonSim}: $\log_{10}A(E)$, $\log_{10}\lambda(E)$, and $\beta(E)$ are each parametrized as cubic polynomials in $\log_{10}E$. This expression refines the first-order estimate $t_0 = s / c$, with $c$ the speed of light, by accounting for the slowdown of particles as they traverse the medium. Because the rate of energy loss depends on both particle type and material, each surrogate light-emission model must include a corresponding $f^{t_0}_\gamma$.

\begin{figure}[htbp]
\centering
\caption{An example muon event generated with \textsc{PhotonSim}. The particle origin is denoted with a red star symbol and the photons creation time is indicated in color. A total of 1k photons chosen at random are also represented by arrows.}
\label{fig:photonsim_muon_event_display}
\includegraphics[width=0.48\textwidth]{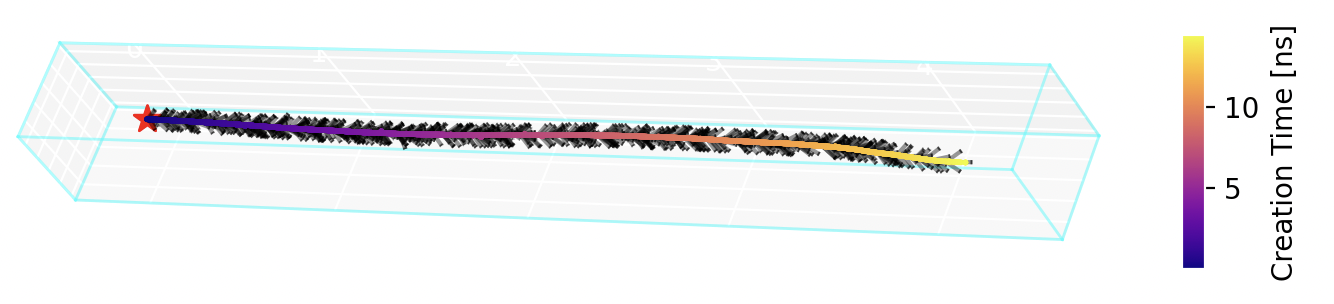}
\end{figure}

\begin{figure}[htbp]
\centering
\caption{Three example energy slices for the surrogate model of $f^{\mathrm{Cherenkov}}$ for muons (top) and electrons (bottom).}
\label{fig:physics_cherenkov_siren}
\includegraphics[width=0.48\textwidth]{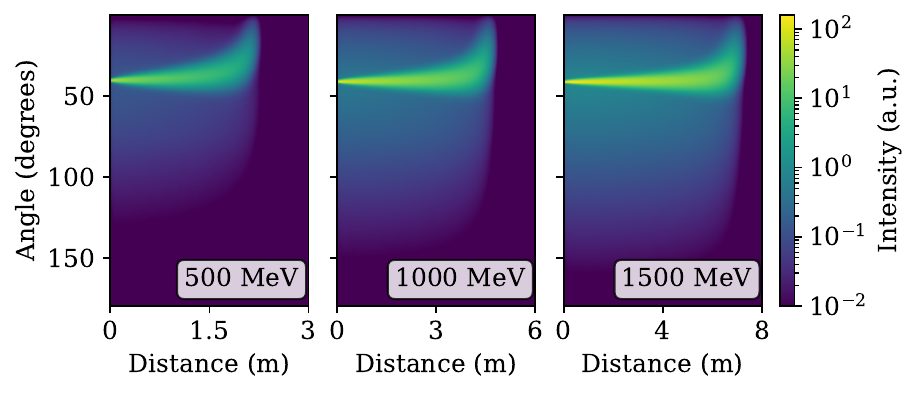}
\includegraphics[width=0.48\textwidth]{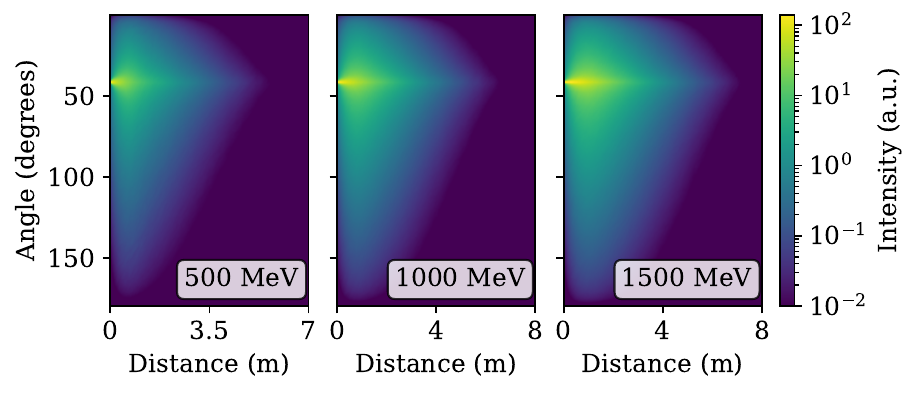}
\end{figure}

\begin{figure}[htbp]
\centering
\caption{Example of the parametrization of $f^{t_0}_\gamma$ for muons in water after subtracting $t_0 = s/c$, which accounts for the $t_0$ correction arising from the particle slowdown in the medium. Data points are shown as circular markers colored by particle energy (see colorbar), with lines of matching colors representing the predictions from Eq.~\ref{eq:t0_correction}.}

\label{fig:t0_correction}
\includegraphics[width=0.44\textwidth]{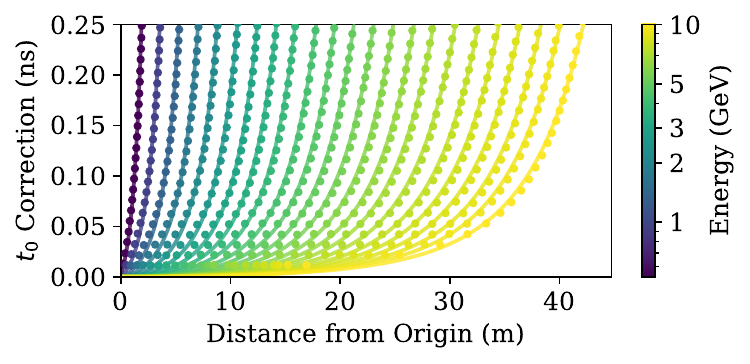}
\end{figure}

An example muon generated with \textsc{PhotonSim} is presented in Figure~\ref{fig:photonsim_muon_event_display}. Examples of $f^{\mathrm{Cherenkov}}_\gamma$ and $f^{t_0}_\gamma$ for different energies are shown in Fig.~\ref{fig:physics_cherenkov_siren} and Fig.~\ref{fig:t0_correction}.

The ray generation function for a particle can be generalized as a sum over all light-producing mechanisms:
\begin{equation}
f^{\rm Particle}_{\gamma} = f^{\mathrm{Cherenkov}}_\gamma + f^{\mathrm{Scintillation}}_\gamma.
\end{equation}
A strategy to model the emission of Cherenkov light along a particle path in a way that allows obtaining gradients for the particle parameters has been outlined above; an analogous strategy for scintillation light is presented in Appendix~\ref{app:scintillation}. In the main body of this manuscript, however, we focus on the Cherenkov-only case for simplicity.

The description of light emission by calibration sources, $f^{\mathrm{Calibration}}_\gamma$, does not require parameterization as a function of a particle path, making it conceptually simpler. The calibration sources used in this work, namely an isotropic point source and a laser (see Sec.~\ref{sec:calibration}), use equal intensities for all rays.

For the isotropic point source, all rays share the same origin and are distributed using a deterministic Fibonacci lattice~\cite{Hannay_2004} rather than pseudo-random sampling. The Fibonacci lattice places $N$ points on the unit sphere according to
\begin{align}
z_i &= 1 - \frac{2i + 1}{N}, \\
\theta_i &= 2\pi \left( i \cdot \varphi \right) \mod 2\pi,
\end{align}
where $i \in \{0, 1, \ldots, N-1\}$ and $\varphi = (\sqrt{5} - 1)/2 \approx 0.618$ is the golden ratio conjugate. The Cartesian coordinates follow as $x_i = \sqrt{1 - z_i^2} \cos\theta_i$, $y_i = \sqrt{1 - z_i^2} \sin\theta_i$. Unlike pseudo-random sampling, which suffers from non-uniform clustering, this quasi-Monte Carlo distribution provides highly uniform coverage of the sphere, significantly reducing the variance associated with finite sample sizes.

For the laser source, rays originate from the fiber tip position and are distributed uniformly in solid angle within a cone. The maximum emission angle is determined by the fiber numerical aperture (NA) and the refractive index of the medium:
\begin{equation}
\theta_{\mathrm{max}} = \arcsin\left( \frac{\mathrm{NA}}{n} \right).
\end{equation}
To achieve uniform sampling in solid angle (rather than uniform in $\theta$), we draw $u \sim \mathcal{U}[0,1]$ and compute
\begin{equation}
\theta = \arcsin\left( \sqrt{u} \sin\theta_{\mathrm{max}} \right),
\end{equation}
with the azimuthal angle $\phi$ sampled uniformly in $[0, 2\pi)$.

More sophisticated calibration sources could be implemented if needed, for example by modeling finite-sized or spatially non-uniform emitters, or by assigning position- or angle-dependent intensities to the rays.

\subsubsection{Wavelength Dependence}
\label{sec:wavelength}

This work treats photon transport and detection as wavelength dependent. Each ray carries its own wavelength $\lambda_i$: fixed to a single value for a monochromatic calibration laser, or, for Cherenkov light, sampled from the $\propto 1/\lambda^{2}$ spectrum within the photosensor-sensitive band ($275$--$674$~nm). This wavelength sets the ray's scattering length, its absorption length, and its detection efficiency. The scattering length combines Rayleigh scattering with a forward Mie component (Appendix~\ref{app:mie}). Both lengths follow the Super-Kamiokande calibration parametrization~\cite{Abe:2013gga}. The detection efficiency uses the measured Hamamatsu R3600 quantum-efficiency curve~\cite{Okajima:2015qoq}, evaluated at the ray's own wavelength when its charge is deposited (Eq.~\ref{eq:hit_charge}). Weighting that curve by the Cherenkov spectrum over the $294$--$648$~nm range in which it is defined gives the mean acceptance $\langle\mathrm{QE}\rangle_C = \int\mathrm{QE}(\lambda)\,\lambda^{-2}\,d\lambda\,/\! \int\lambda^{-2}\,d\lambda \approx 12.8\%$, which characterizes the photosensor response to Cherenkov light. Rays generated outside that range are propagated but deposit no charge. The one exception is the Cherenkov emission surrogate $f^{\mathrm{Cherenkov}}_\gamma$, which is trained on the wavelength-integrated photon yield. The emission angle and the photon wavelength are therefore independent. The wavelength is drawn separately at generation. Several extensions remain for future work: an explicit wavelength axis for the emission surrogate, which would couple emission angle and wavelength; a wavelength-dependent refractive index and surface reflectivity, which are achromatic here; and a model of photon polarization.

\subsection{Ray Propagation}
\label{sec:photon_prop}

We propagate rays through the detector medium using an analytic ray tracing approach. Unlike ray marching methods that discretize trajectories into small steps, our approach calculates exact intersection points with the detector geometry analytically. This ensures precision is limited only by floating-point arithmetic. The computational cost is then set by the fixed number of interactions $K$ a ray is allowed, rather than by the distance it travels.

The propagation process is iterative. Each ray carries a position and a direction, an intensity $I$ fixed at generation, and a survival factor $\alpha$. The survival factor accumulates the probability that the ray has survived the interactions encountered so far. At each step $k$, up to a maximum $K$, we first locate the ray collision point (Sec.~\ref{sec:grid_accel}). We then evaluate the probabilities of the competing physical processes: scattering, absorption, reflection and detection. The expected detected signal is deposited and the surviving ray is attenuated. Finally the ray is advanced to its next state by differentiable sampling. The following subsections expand on these steps in more detail.

\subsubsection{Grid-accelerated Intersection}
\label{sec:grid_accel}

A naive intersection algorithm checks every ray against every photosensor, scaling linearly with the number of sensors $O(N_s)$. For existing large-scale experiments, involving thousands of photosensors, this is prohibitively expensive. We overcome this by implementing a spatially hashed grid system that reduces the complexity to $O(1)$.

\begin{figure}
\centering
\caption{Visualization of the grid-based sensor lookup on the top cap of an SK-like detector. A single active grid cell is mapped to a specific subset of assigned sensors, allowing the intersection algorithm to ignore the remaining inactive sensors.}
\label{fig:grid_cell}
\includegraphics[width=0.44\textwidth]{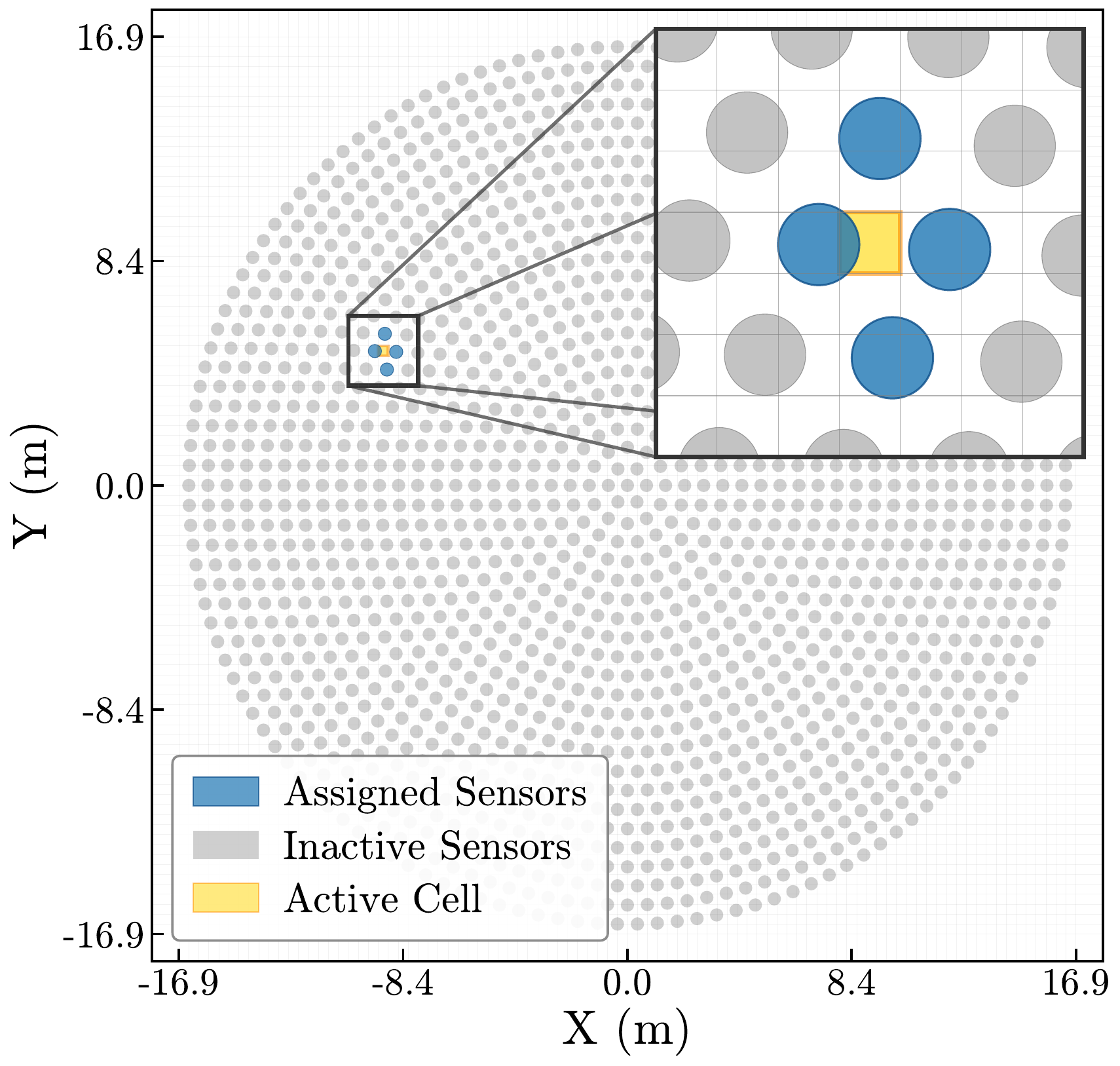}
\end{figure}

For a given detector geometry, we discretize the surface into a grid of cells. During initialization, we precompute a mapping that assigns each grid cell a fixed list of up to $M$ nearby sensors (typically $M=4$), prioritized by geometric overlap and proximity. We model the photosensor enclosures as spheres to simplify these overlap calculations. During propagation, we first compute the analytic intersection of the ray with the global detector volume to obtain the surface intersection coordinates. These coordinates map directly to a specific grid cell. The mapping is an arithmetic index computation rather than a search, so the lookup is $O(1)$. The number of cells therefore sets the memory of the acceleration structure, not the per-ray cost, which is bounded by $M$ and independent of the total sensor count. The cell returns a candidate list of sensors that the ray may strike. This efficient reduction of the search space is illustrated in Fig.~\ref{fig:grid_cell} for the top cap of a cylindrical detector. This two-stage approach yields the collision distance $d$, the surface normal $\mathbf{n}$, and the indices of the specific sensors contributing to the hit.

\subsubsection{Probabilistic Interactions and Variance Reduction}

Given a ray traveling a distance $d$ to a surface, we model the medium interactions using the scattering length $\lambda_s$ and absorption length $\lambda_a$. Both are evaluated at the ray's wavelength; the explicit $\lambda$ argument is suppressed in the expressions below. The probability that a photon reaches the surface without scattering is $P_{\mathrm{reach}} = \exp(-d/\lambda_s)$, while the probability of scattering is $P_{\mathrm{scatter}} = 1 - P_{\mathrm{reach}}$.

Standard Monte Carlo methods sample a binary outcome (scatter or reach) at this stage. To reduce variance and ensure smoother gradients, we employ \emph{implicit capture}. Instead of terminating rays probabilistically, we deposit the expected probability weight directly into the sensor and attenuate the surviving ray. The ray carries an intensity $I_i$ and a survival factor $\alpha_i$. We calculate the attenuation due to absorption as $A(d) = \exp(-d/\lambda_a)$.

If the ray reaches the surface, it may reflect with probability $R$ or be detected with probability $P_{\mathrm{detect}} = P_{\mathrm{reach}} \cdot (1 - R)$. We deposit a fraction of the ray intensity into the sensor:\begin{equation}n_{\rm dep} = I_i \cdot \alpha_i \cdot P_{\mathrm{detect}} \cdot A(d) \cdot w_{\rm geo},\end{equation}where $w_{\rm geo}$ is a geometric overlap factor described in Sec.~\ref{sec:photon_relaxation}. The ray then continues to propagate with its intensity scaled by the probability of the surviving paths (reflection or scattering) and their respective attenuation factors.

The factorization of the ray signal into a generation intensity $I_i$, set once at emission, and a survival factor $\alpha_i$, updated at each propagation step, is also motivated by gradient stability. Because $\alpha_i$ decays exponentially with path length, combining it with $I_i$ into a single weight would cause the reverse-mode chain rule to divide by this exponentially small quantity when computing gradients with respect to generation-stage parameters, leading to numerical blow-up. Maintaining the two factors as separate nodes in the computation graph reorganizes the backward pass so that gradients of the generation intensity bypass the survival attenuation path entirely, keeping them well-conditioned throughout the propagation loop.

\subsubsection{Photon Relaxation}
\label{sec:photon_relaxation}
A fundamental challenge in differentiable ray tracing is the binary nature of geometric intersection. Typically, the contribution of a ray to a sensor is defined by an indicator function which is 1 if the ray hits the sensor and 0 otherwise. This discontinuity prevents gradients from guiding rays toward sensors during optimization, as the gradient is zero almost everywhere.

\begin{figure*}[htbp]
\centering
\caption{\textbf{(Left)} Geometric definition of the photon relaxation model. A photon with Gaussian width $\sigma$ passes a sensor of radius $r$ at an impact parameter $b$. The purple shaded region illustrates the integral of the photon's probability density over the sensor surface, which determines the weight $w_{\rm geo}$ assigned to the hit. \textbf{(Right)} The calculated overlap probability $w_{\rm geo}$ as a function of the normalized impact parameter $b/r$. As the smoothing width $\sigma$ increases, the response transitions from a non-differentiable step function (binary) to a smooth curve that provides useful gradients even when the ray strictly misses the sensor ($b/r > 1$).}
\label{fig:overlap_relaxation}
\includegraphics[width=0.45\textwidth]{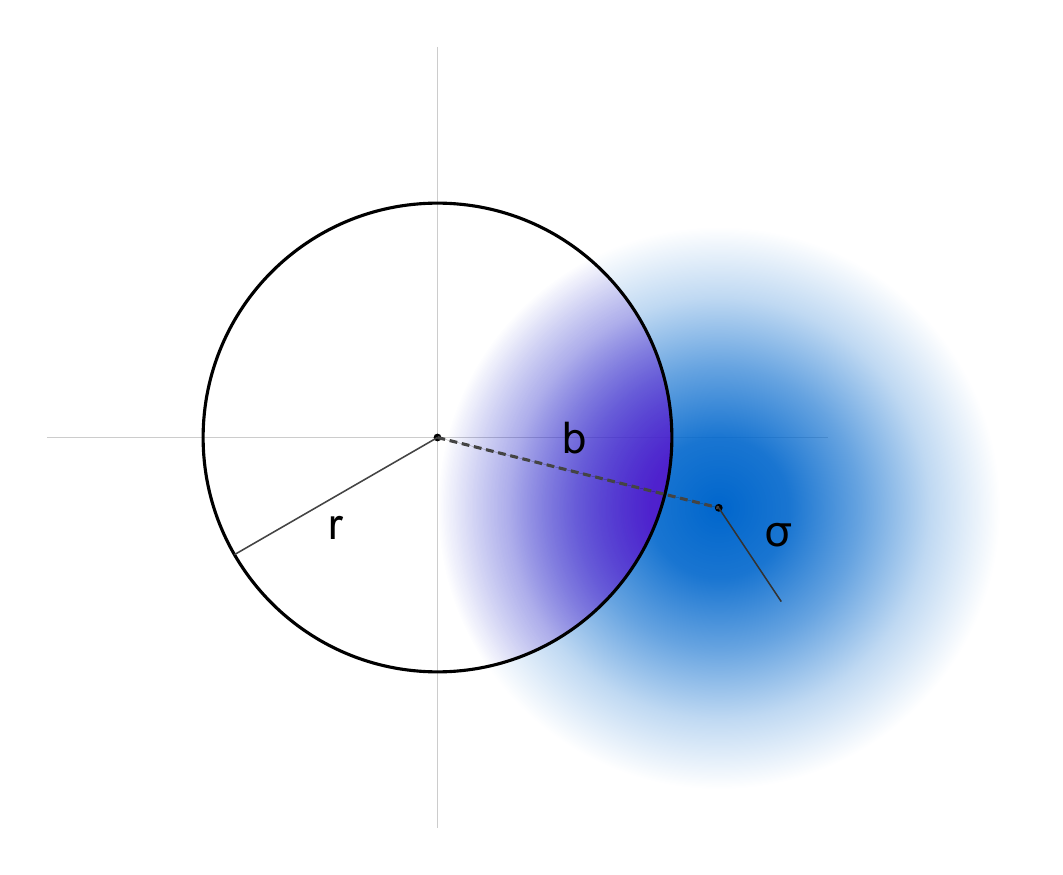}
\includegraphics[width=0.45\textwidth]{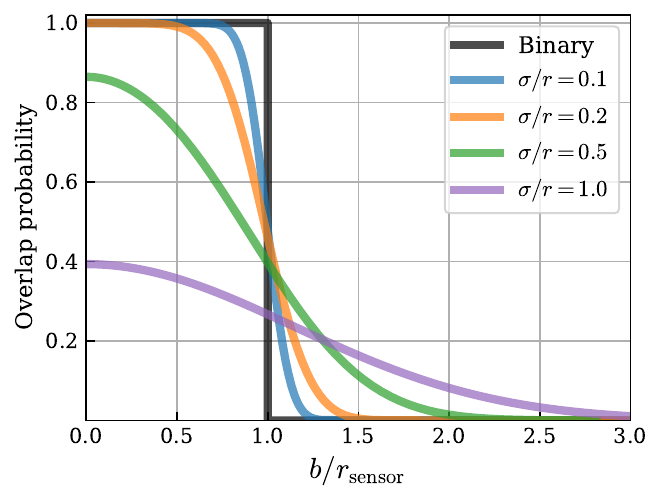}
\end{figure*}

Classic smooth relaxations, such as applying a sigmoid function to the signed distance, can address the discontinuity but often distort the effective sensor shape and underlying physics. We address this via \emph{photon relaxation}, where we treat photons as having a spatial extent modeled by a Gaussian probability density function with width $\sigma$ perpendicular to the propagation direction. The overlap weight $w_{\rm geo}$ corresponds to the integral of this Gaussian over the circular sensor area, represented by the purple region in Fig.~\ref{fig:overlap_relaxation}. Exploiting the rotational symmetry, we precompute this integral as a function of the single impact parameter $b$, defined as the perpendicular distance from the ray to the sensor center, and store it as a differentiable lookup table. This formulation ensures that rays produce a nonzero, distance-dependent signal even when they strictly miss a sensor ($b > r$). In the tracking studies $\sigma$ is one tenth of the sensor radius. The calibration fits retain the strict intersection and apply no relaxation. Further details on the numerical implementation are provided in Appendix~\ref{app:overlap}.

\subsubsection{Differentiable Path Sampling}

Having deposited the expected charge, the ray must be advanced to its next state. It either scatters in the volume before reaching a surface or reflects there, specularly or diffusely. A new free path and outgoing direction are then drawn. These outcomes are random, so differentiating a Monte Carlo simulation with respect to physical parameters is not automatic. A derivative computed along one sampled path captures how that path's deposits change, but it is blind to the fact that the parameter also changes which paths are sampled in the first place. Nudging the scattering length alters not only the charge each ray deposits but the probability that it scattered at all. This second effect leaves no trace in the arithmetic of any single fixed path. Parameters enter the expected response in three ways, only two of which require a dedicated gradient technique. Some set the amount of detected light (the absorption length, the reflectivities, the quantum efficiency) or scale the emission (the particle energy and photon yield); these enter the deposited charge directly and ordinary automatic differentiation handles them. Others set the value of a continuous random draw (the free path, the scattering and reflection angles); these we treat by reparameterization. A third kind sets the probability of a choice: the scattering length, which decides whether a ray scatters before reaching the surface, and the specular fraction, which decides the reflection branch; these we treat with the score function.

The distinction is made precise by a single identity for the derivative of an expected signal. Let $\omega$ denote the collection of random outcomes that defines one ray's path: whether and where it scattered, the sampled distances and angles, and the reflection branch taken at each surface. Let $\Phi(\omega;\theta)$ be the signal that path deposits, the charge $n_{\mathrm{dep}}$ of the preceding section, let $p(\omega;\theta)$ be the probability of realizing that particular path under the current value of a physical parameter $\theta$, and let $\mathbb{E}_{\omega}$ denote the average over random paths, that is, over the simulated rays. Then
\begin{equation}
\label{eq:score}
\frac{\partial}{\partial\theta}\,\mathbb{E}_{\omega}\!\left[\Phi(\omega;\theta)\right]
= \mathbb{E}_{\omega}\!\Big[\,
\underbrace{\tfrac{\partial \Phi}{\partial\theta}}_{\text{direct}}
+ \underbrace{\Phi\,\tfrac{\partial \log p(\omega;\theta)}{\partial\theta}}_{\text{score}}\,\Big].
\end{equation}
The direct term is the change in the charge deposited along the path when $\theta$ varies with the path held fixed. The score term is the change in the expectation that arises because $\theta$ makes some paths more or less probable: the score $\partial \log p/\partial\theta$ is large where the parameter strongly reweights a choice's probability, and it enters multiplied by the charge $\Phi$ that the path deposits. The intuition is that raising a parameter that makes a high-charge outcome more likely raises the expected charge, even though no individual deposit changes value. The amount and scaling parameters listed above contribute only to the direct term, and the specular fraction only to the score term. The scattering length contributes to both: it sets the probability that a ray scatters before the surface, and it also enters the deposited charge through the reach factor $\exp(-d/\lambda_s)$.

For the direct term to cover the continuous draws, the sampled values themselves must be differentiable functions of the parameters. This is arranged with the \emph{reparameterization trick}~\cite{kingma_2013}: each draw is written as a fixed, parameter-free uniform noise sample pushed through a deterministic, parameter-dependent transform. The randomness is thereby separated from the physics; with the noise frozen, the sampled free path or angle, and the ray position and arrival time computed from it, vary smoothly with the parameters like any other intermediate quantity. For volumetric Rayleigh scattering, the usual rejection sampling is not differentiable, so we use inverse-transform sampling instead. Equating the Rayleigh cumulative distribution to the uniform noise yields a depressed cubic for the scattering-angle cosine, which Cardano's formula solves in closed form. The result is a differentiable mapping from noise to angle. The scattering distance is treated the same way. It is drawn by inverting the exponential attenuation truncated at the distance to the boundary, so the scattering point is the ray origin advanced by that length. Appendix~\ref{app:reparam} gives both mappings in full.

At a boundary, each surface reflects as a mixture of a specular and a diffuse (Lambertian) component, described by a reflectivity and a specular fraction: $(R_w,f_w)$ for the detector walls and $(R_s,f_s)$ for the sensor surfaces. The idealization of purely specular sensors and purely diffuse walls is recovered in the limit $f_s=1$, $f_w=0$. The specular direction is a deterministic function of the incident direction and the surface normal, so it is differentiable as it stands, and the diffuse direction is reparameterized from cosine-weighted hemispherical sampling (Appendix~\ref{app:reparam}). The choice between the two branches, however, is a discrete outcome whose probability is the specular fraction itself. Its gradient is therefore carried by the second term of Eq.~\ref{eq:score}.

In practice we evaluate the score term with the Infinitely Differentiable Monte Carlo Estimator (DiCE)~\cite{foerster2018dice}. As a ray propagates, it accumulates a running log-score $\ell$, the sum of the log-probabilities of the choices it has realized. Every deposited charge is then multiplied by the factor $\exp\!\left(\ell-\mathrm{sg}(\ell)\right)$. The stop-gradient operator $\mathrm{sg}(\cdot)$ returns its argument in the forward pass and has zero derivative in the backward pass. The factor is therefore exactly unity in the forward pass and changes no simulated value, but differentiating the product $n_{\mathrm{dep}}\exp\!\left(\ell-\mathrm{sg}(\ell)\right)$ produces $\partial n_{\mathrm{dep}}/\partial\theta + n_{\mathrm{dep}}\,\partial\ell/\partial\theta$, precisely the direct and score terms of Eq.~\ref{eq:score} for every deposit downstream of a stochastic choice. Unlike the elementary likelihood-ratio estimator, this construction remains correct under repeated differentiation, so it composes safely with the rest of the automatic-differentiation machinery; the same differentiated pass supplies the first-derivative Jacobian that is assembled into the Fisher metric of Sec.~\ref{sec:gauss_newton}.

With the next state selected, the ray advances to the following iteration ($k \rightarrow k+1$): the global time is incremented by the flight time over the sampled path length, and the survival factor $\alpha$ is scaled by the absorption along that path. This cycle of intersection, deposition, and state update repeats for a fixed number of iterations $K$, chosen so that the unsimulated residual intensity is negligible (Appendix~\ref{app:Kconvergence}). Gradients propagate through the full depth of the loop. The only term we drop is the derivative of the surface normal, the local curvature, which compounds across bounces and destabilizes the deep gradient. The normal's value still drives the reflected direction and the reflectance, and both remain differentiable through the incident ray (Appendix~\ref{app:grad_truncation}). The complete propagation logic is given in Appendix~\ref{app:propagation}.

\subsection{Hit Making}
\label{sec:hit_making}
The final step converts the predicted rays arriving at each sensor into continuous, differentiable observables that feed the calibration and reconstruction studies of Secs.~\ref{sec:calibration} and~\ref{sec:tracking}. The expected charge at sensor $j$ is the sum of the deposited ray contributions across all propagation steps, each weighted by the wavelength-dependent quantum efficiency $\mathrm{QE}(\lambda)$ of the photosensor (Sec.~\ref{sec:wavelength}):
\begin{equation}
\label{eq:hit_charge}
\mu_j = \sum_{(i, k) \in \mathcal{H}_j} \mathrm{QE}(\lambda_i)\,
n_{\mathrm{dep}, i, k},
\end{equation}
where $\mathcal{H}_j$ is the set of ray--step pairs $(i, k)$ intersecting sensor $j$, $n_{\mathrm{dep}, i, k}$ is the fractional photon count deposited by ray $i$ at step $k$, and $\lambda_i$ its wavelength. For timing, the observable of interest is the arrival time of the \emph{first} detected photon at each sensor. This is an order statistic, so reducing the rays to a single representative time, such as the earliest sampled arrival, would introduce a dependence on the number of simulated rays. Instead, the hit-making step retains the predicted per-ray arrival times $\{t_i\}$ and intensities $w_i = \mathrm{QE}(\lambda_i)\,n_{\mathrm{dep},i}$ reaching each sensor. From these it forms the expected cumulative count of detected photons before a time $t$,
\begin{equation}
R_j(t) = \sum_{i \in \mathcal{H}_j} w_i\,\tfrac{1}{2}\!\left[
1 + \mathrm{erf}\!\left(\frac{t - t_i}{\sigma\sqrt{2}}\right)\right],
\end{equation}
where $\sigma$ is the per-photon time resolution (the photosensor transit-time spread). Being a sum of smooth terms, $R_j(t)$ is an exact, differentiable expectation that converges with the number of rays, unlike a hit-level time average. From it we obtain the survival probability $S_j(t) = 1 - R_j(t)/\mu_j$, the probability that a detected photon arrives after $t$. The distribution of the earliest arrival follows from it directly. This construction supplies the timing term of the reconstruction loss (Sec.~\ref{sec:losses}). Because $\sigma$ is matched to the transit-time spread used when generating the data, it provides a consistent first-arrival model between simulation and reconstruction.

\subsection{Performance}
\label{sec:performance}

\begin{figure}[htbp]
\centering
\caption{Computation times for an SK-like detector geometry, shown as a function of the number of rays ($N_{\mathrm{rays}}$) and the number of propagation iterations ($K$). Lines are linear fits presented in the legend.}
\label{fig:comp_performance}
\includegraphics[width=0.49\textwidth]{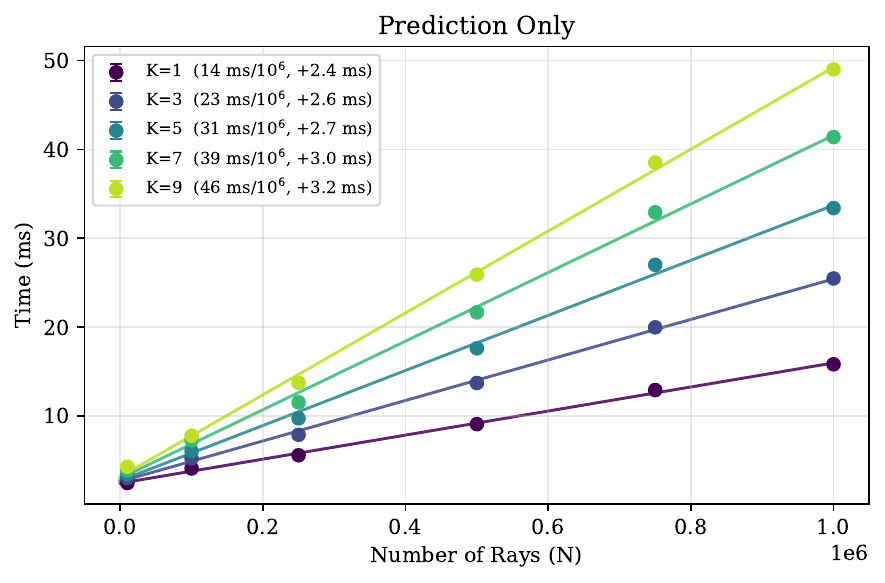}
\includegraphics[width=0.49\textwidth]{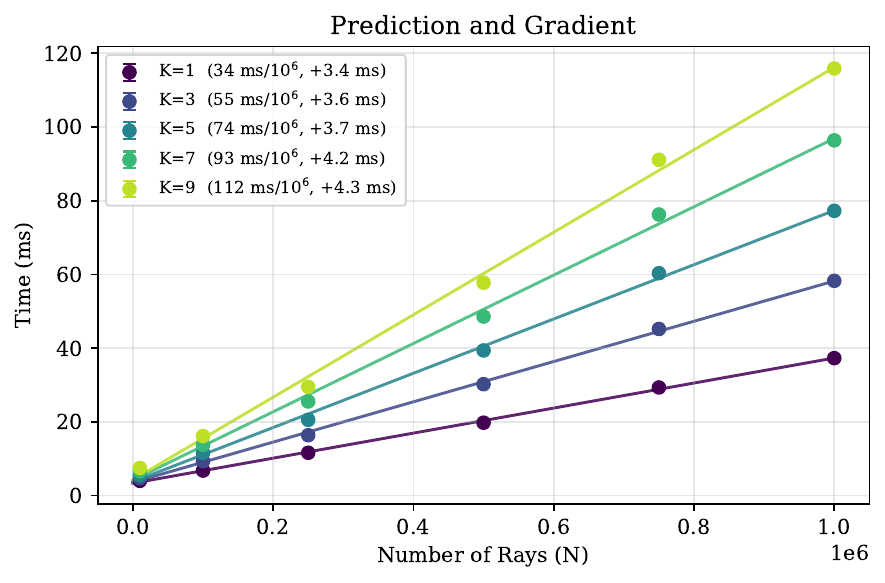}
\includegraphics[width=0.49\textwidth]{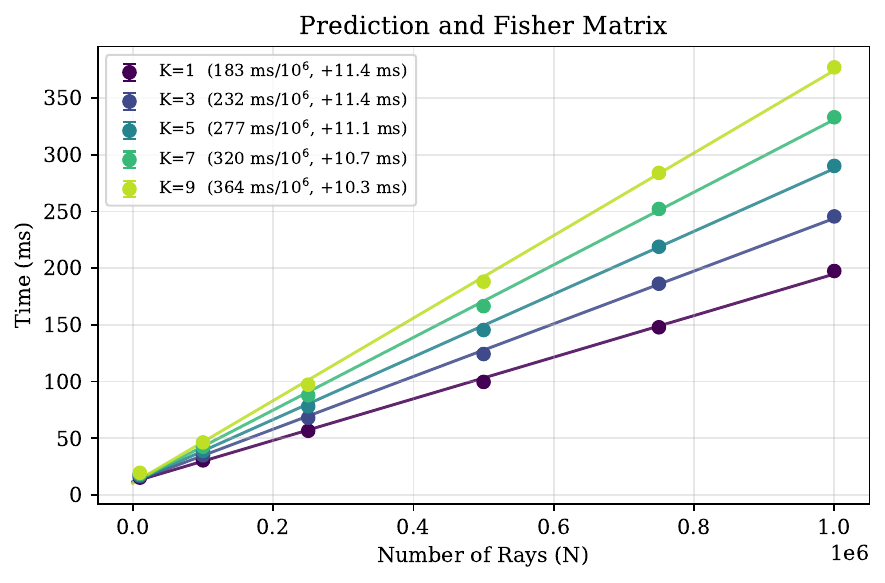}
\end{figure}

A key practical requirement for end-to-end differentiable simulation is that derivative evaluation must not introduce prohibitive overhead. In \lucid, the dominant cost is ray propagation, which grows with the number of simulated rays \(N_{\mathrm{rays}}\) and the maximum number of propagation iterations \(K\). Figure~\ref{fig:comp_performance} quantifies this scaling in an SK-like geometry using a single NVIDIA A100 GPU for the three quantities the fits require: the forward prediction of the expected hit pattern (``Prediction Only''), the prediction together with the reverse-mode gradient of the scalar loss of Eq.~\ref{eq:total_loss} (``Prediction and Gradient''), and the prediction together with the Fisher matrix of Eq.~\ref{eq:fisher}, assembled in forward mode as described in Sec.~\ref{sec:gauss_newton} (``Prediction and Fisher Matrix''). The derivative modes necessarily include the forward pass: reverse-mode AD must evaluate the prediction to build the computation graph, and the forward-mode Jacobian carries it alongside the tangents.

Across the full scanned range, the runtime increases smoothly with both \(N_{\mathrm{rays}}\) and \(K\), and is linear in \(N_{\mathrm{rays}}\) at fixed \(K\). For the tracking working point (\(N_{\mathrm{rays}}\sim10^{5}\) and \(K=8\)), prediction alone executes in \(\mathcal{O}(10)\,\mathrm{ms}\). The calibration fits run at \(10^{6}\) rays. The ray budget is a free choice traded against the available compute and the precision required, and the values used throughout are collected in Appendix~\ref{app:optimizer}. For comparison, the average number of Cherenkov photons produced in a single event is $\approx 2\times10^{5}$ for a 1~GeV muon, within the SK-like photosensor-sensitive band (275--674~nm), and scales almost linearly with the particle energy. Nonetheless, rays do not behave as individual photons: each deposits an expectation-weighted intensity across multiple sensors, so matching the number of physical photons is not necessarily required.

The dependence on \(K\) is linear but not proportional. Fitting the measured slope to \(a+bK\) leaves a sizeable \(K\)-independent term, so going from \(K=1\) to \(K=9\) costs a factor \(\sim\!3.5\) rather than nine. That offset is the per-event work which does not repeat with depth, namely ray generation and per-sensor accumulation, while the \(K\)-dependent part measures the per-step propagation logic.

Adding the gradient costs a factor \(\sim\!2\) over prediction alone, roughly constant across the scanned range. This behavior is a direct consequence of reverse-mode AD: once the forward computation graph is constructed, the cost of obtaining all derivatives of a scalar loss with respect to a (potentially large) parameter vector is roughly a constant factor times the forward cost, rather than scaling linearly with the number of optimized parameters. Adding the Fisher matrix instead costs a factor \(\sim\!8\) in the median, since it carries one forward-mode tangent per fitted parameter; this ratio falls from \(\sim\!12\) at \(K=1\) to \(\sim\!5\) at \(K=9\) as the per-ray work amortizes them. This scalar-loss gradient underlies the gradient-based phases of the fits that follow. The Gauss--Newton calibration and tracking of Secs.~\ref{sec:calibration} and~\ref{sec:tracking} instead assemble the Jacobian $\partial\bm\mu/\partial\bm\theta$ in forward mode over a small parameter vector. The $\sim\!10^{4}$ per-sensor efficiencies add no columns to it, because they are written as a function of the optical parameters rather than fitted alongside them (Sec.~\ref{sec:gauss_newton}). Either way the derivative cost is a small multiple of one forward simulation, independent of the number of parameters inferred.

Taken together, these timing results demonstrate that \lucid supports gradient-based calibration and reconstruction at a computational scale within an order of magnitude of forward simulation, with runtime controlled primarily by user-chosen accuracy knobs \((N_{\mathrm{rays}}, K)\) rather than by the dimensionality of the inference problem.

\subsection{Generating Data-like Events}
\label{sec:data_like}

To validate and benchmark the differentiable simulation, we generate \emph{data-like} events, i.e., fully stochastic realizations that reproduce the statistical characteristics of real detector data for a given geometry and configuration. These events serve as reference samples against which to compare the differentiable, expectation-based predictions described in previous sections.

The light comes directly from \textsc{PhotonSim} rather than the SIREN surrogate. Each event injects a GEANT4 photon-by-photon prediction (Fig.~\ref{fig:photonsim_muon_event_display}), so the microscopic emission is sampled in full.

Propagation uses the same optical processes as Sec.~\ref{sec:photon_prop}, but every process is drawn as a discrete event rather than averaged into a continuous weight. The photon-relaxation model is disabled. Every photon then follows its own random trajectory, reproducing the stochasticity of real photon transport.

Each propagated photon carries unit weight and is tested for detection individually. Detection is a per-photon Bernoulli trial whose success probability is the wavelength-dependent quantum efficiency $\mathrm{QE}(\lambda)$ of the photosensor, evaluated at the photon wavelength from the measured Super-Kamiokande PMT quantum efficiency response. The number of detected photoelectrons at each sensor is the count of its photons that pass this trial. The recorded charge is then obtained by smearing this photoelectron count to reproduce the photosensor charge resolution, following the Super-Kamiokande calibration~\cite{Abe:2013gga}: a zero-mean Gaussian whose fractional width depends on the collected charge $Q$, with $\sigma/Q \simeq 1.2\%$ below $20$~photoelectrons, $0.75\%$ up to $130$, and $0.5\%$ above. The smearing preserves the mean, so the expected recorded charge equals the photoelectron count, while reproducing the physical charge resolution of the detector.

Timing is handled at the photon level. Each photon arrival time is smeared independently by a Gaussian of width $\sigma_{\mathrm{TTS}} = 2.1$~ns, the photosensor transit-time spread (TTS), matching the single-photoelectron timing resolution of the Super-Kamiokande PMTs~\cite{Abe:2013gga}. The sensor hit time is then the earliest arrival among all detected photons, computed as an exact minimum, in contrast to the smooth survival-based estimator used for the differentiable expectation (Sec.~\ref{sec:hit_making}). Finally, a per-event $t_0$ offset drawn uniformly in a $\pm 15$~ns window is added to all photon times before hit formation.

The Super-Kamiokande dark rate of $\approx4.2$~kHz per PMT~\cite{Abe:2013gga} amounts to a handful of dark hits across the full array in a typical event window of $\mathcal{O}(100)$~ns. Their precise impact depends on the trigger window assumed, and matters for low-energy physics; we neglect them here, since the $\mathcal{O}(\mathrm{GeV})$ particles considered produce orders of magnitude more hits.

\begin{figure}[htbp]
\centering
\caption{Example of a 1.5~GeV muon data-like event in the SK-like detector. Dark noise is not included.}
\label{fig:data_like_track}
\includegraphics[width=0.49\textwidth]{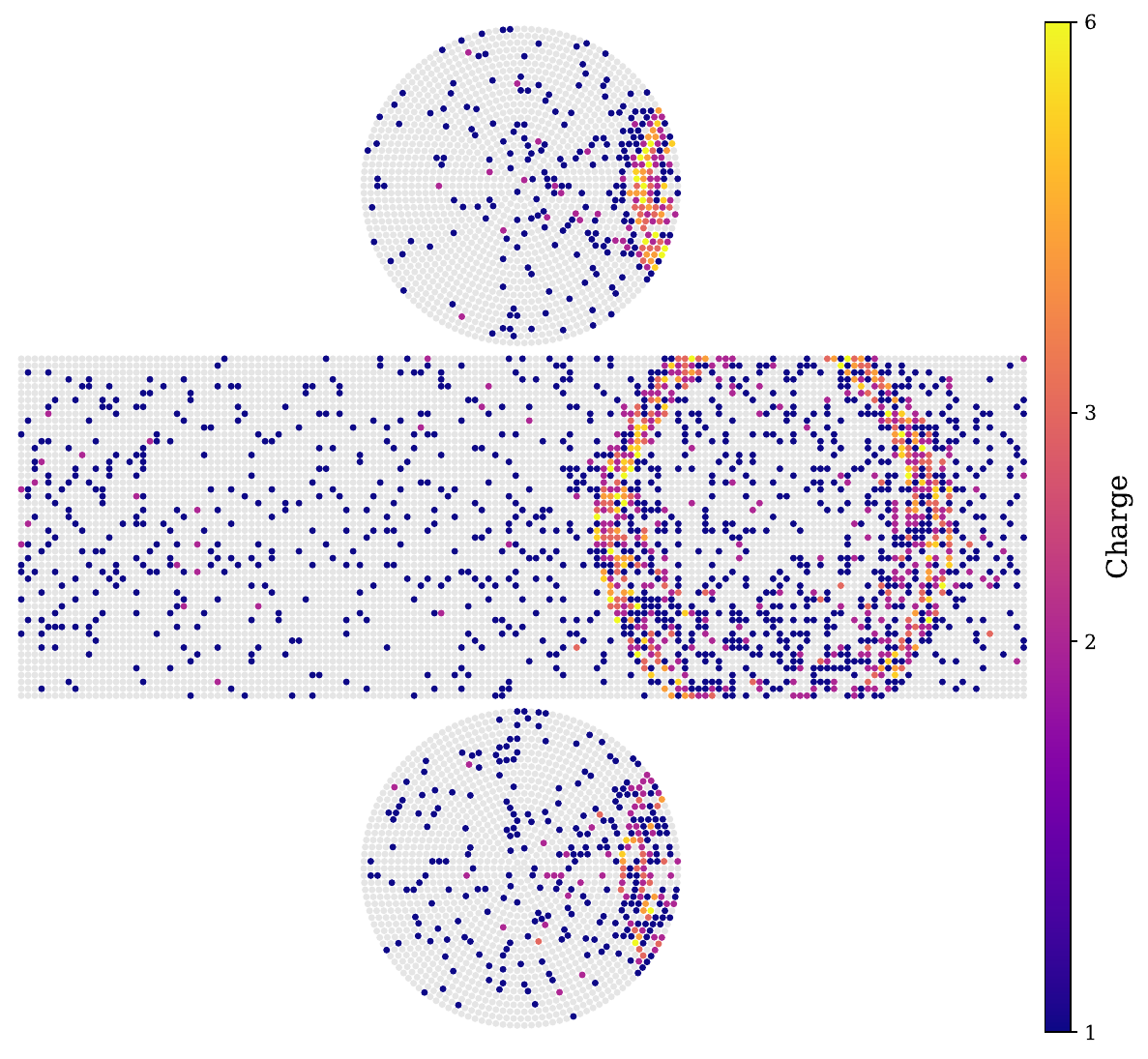}
\includegraphics[width=0.49\textwidth]{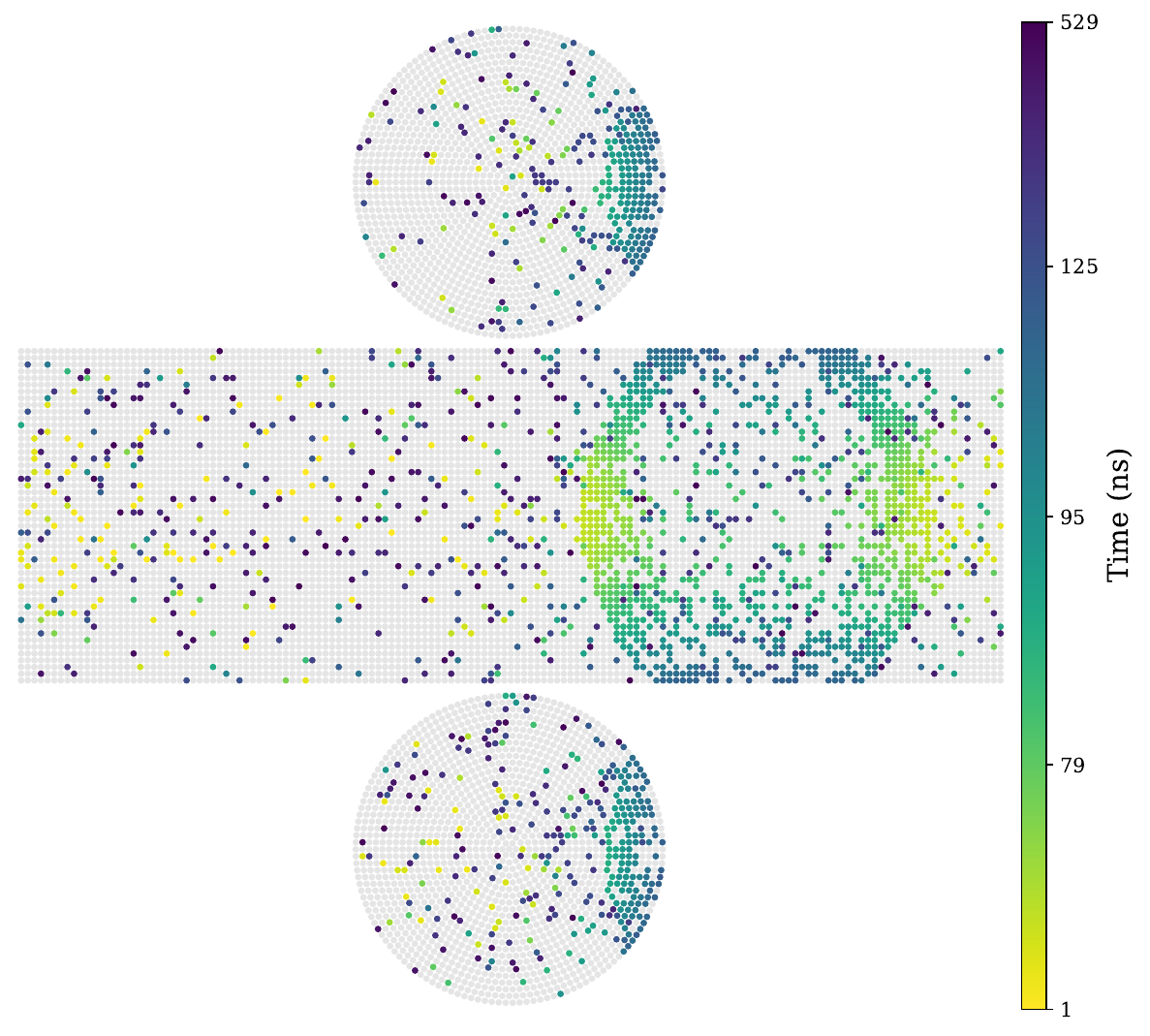}
\end{figure}

Figure~\ref{fig:data_like_track} shows the sampled charge and time for a data-like muon event in the SK-like detector.

\begin{figure}[htbp]
\centering
\caption{Example data-like events and associated differentiable predictions for various detector geometries. The color intensity represents the observed counts in logarithmic scale.}
\label{fig:pred_vs_data}
\includegraphics[width=0.48\textwidth]{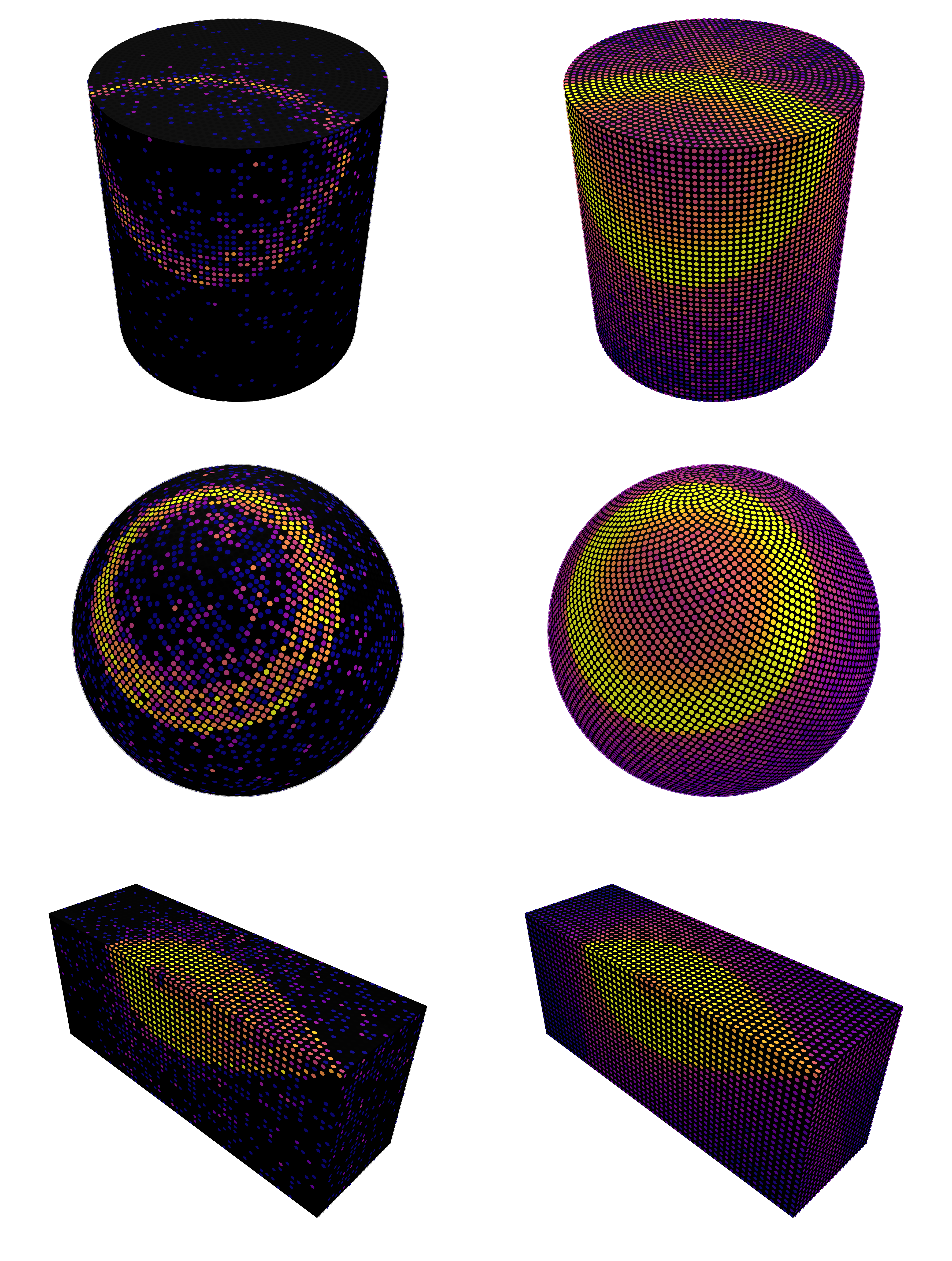}
\end{figure}

Figure~\ref{fig:pred_vs_data} places data-like events beside their differentiable expectations for several geometries. Both are generated with identical particle and detector configurations.

\subsection{Losses}
\label{sec:losses}

To illustrate gradient-based calibration and optimization, we employ the loss functions described below. Both calibration and reconstruction compare the per-sensor observables produced by the hit-making step with those of the data. These observables are the hit charge and time, which measure, respectively, the charge collected by the sensor, in photoelectrons, and the arrival time of the earliest detected photoelectron. From them we define two loss terms, a charge term $\mathcal{L}_{\mathrm{counts}}$ and a first-arrival time term $\mathcal{L}_{\mathrm{time}}$.

The charge term is the Poisson negative log-likelihood of the observed charge $q_j$ given the predicted expected charge $\mu_j$ of Eq.~\ref{eq:hit_charge},
\begin{equation}
\label{eq:counts_loss}
\mathcal{L}_{\mathrm{counts}}
= \sum_j \left( \mu_j - q_j \log \mu_j + \log \Gamma(q_j + 1) \right).
\end{equation}
The charge, expressed in photoelectrons, folds the single-photoelectron electronics response applied when generating the data and is treated here as an effective photoelectron count; the $\log \Gamma(q_j + 1)$ term does not depend on the fit parameters and does not contribute to the gradient. Because the total predicted light $\sum_j \mu_j$ grows with the particle energy, this expression is kept unnormalized, so that the absolute charge constrains the energy scale.

The simulator places no restriction on the objective; any differentiable function of the predicted and observed quantities can be minimized. The Poisson likelihood and the least-squares $\chi^2$ statistics are the physically motivated choices for counting data. A standard alternative to Eq.~\ref{eq:counts_loss} is therefore the Neyman $\chi^2$,
\begin{equation}
\label{eq:chi2_loss}
\mathcal{L}_{\chi^2} = \sum_j \frac{(\mu_j - q_j)^2}{q_j},
\end{equation}
in which each residual is weighted by its observed count rather than by the predicted mean. Which of these best suits a given measurement is a modeling choice, and Appendix~\ref{app:optimizer} relates the two forms.

The time term is built from the survival probability $S_j(t) = 1 - R_j(t)/\mu_j$ introduced in Sec.~\ref{sec:hit_making}, the probability that a detected photon reaches sensor $j$ later than $t$. For a hypothesized event start time $t_0$, let $\tilde{t}_j = t_j^{\mathrm{obs}} - t_0$ be the observed hit time of a lit sensor relative to that start time. Treating the observed charge $q_j$ as the effective number of detected photoelectrons and the hit as their earliest arrival, the probability that it falls within a narrow window of width $\delta$ around $\tilde{t}_j$ is $S_j(\tilde{t}_j - \tfrac{\delta}{2})^{q_j} - S_j(\tilde{t}_j + \tfrac{\delta}{2})^{q_j}$, since $S_j(t)^{q_j}$ is the probability that all $q_j$ photoelectrons arrive after $t$. The time loss is the corresponding negative log-likelihood, summed over lit sensors,
\begin{equation}
\label{eq:time_loss}
\mathcal{L}_{\mathrm{time}}
= -\sum_{q_j > 0} \log \left[
S_j\!\left(\tilde{t}_j - \tfrac{\delta}{2}\right)^{q_j}
- S_j\!\left(\tilde{t}_j + \tfrac{\delta}{2}\right)^{q_j} \right].
\end{equation}
Sensors with no detected light do not contribute. Because $R_j$, and hence $S_j$, is a convergent expectation rather than a sampled minimum, this term remains well defined for any number of simulated rays.

The two terms are the negative logarithms of the factorized per-sensor likelihood $P(q_j)\,P(t_j^{\mathrm{obs}} \mid q_j)$, which the reconstruction loss combines as
\begin{equation}
\label{eq:total_loss}
\mathcal{L} = \mathcal{L}_{\mathrm{counts}} + w\,\mathcal{L}_{\mathrm{time}},
\end{equation}
with the relative weight $w$ set to unity by default. To keep the two terms from competing during the optimization, the time term is evaluated with the predicted per-ray intensities and their totals $\mu_j$ both held fixed through a stop-gradient. Its gradient then flows only through the predicted arrival times, constraining the vertex position, direction, and event start time. The charge and energy information is carried by $\mathcal{L}_{\mathrm{counts}}$. This separation of charge and time keeps the joint objective well conditioned.

\subsection{Second-Order Parameter Estimation}
\label{sec:gauss_newton}

Both analyses in this work, calibrating the detector parameters (Sec.~\ref{sec:calibration}) and reconstructing particle tracks (Sec.~\ref{sec:tracking}), reduce to the same task: find the parameters that minimize the per-sensor loss of Sec.~\ref{sec:losses}. Calibration constrains the detector parameters from charge alone, while tracking adds the first-arrival time term $\mathcal{L}_{\mathrm{time}}$ and fits the seven track parameters. Because the simulator is differentiable, we solve both with the same second-order optimizer, described once here and specialized in the two sections.

We write the parameters as $\bm\theta$ and the predicted expected charge at sensor $j$ as $\mu_j(\bm\theta)$ (Eq.~\ref{eq:hit_charge}), with $q_j$ the observed charge. Because each $\mu_j$ is the output of a stochastic ray-tracing simulator, every loss evaluation is a Monte Carlo estimate. The curvature introduced below is accordingly averaged over several independent noise realizations, and the reported estimate is the Polyak average~\cite{polyak1992acceleration} of the final iterates, which suppresses the residual stochastic jitter. The averaging and refresh schedules are given in Appendix~\ref{app:optimizer}.

The gradient $\bm g \equiv \partial\mathcal{L}/\partial\bm\theta$ fixes the locally best direction of improvement but not how far to move: a first-order update $\bm\theta \leftarrow \bm\theta - \eta\,\bm g$ needs an external step size $\eta$, and a single $\eta$ must serve parameters the data constrain very unequally. First-order optimizers such as Adam~\cite{Kingma:2014vow} infer per-parameter step sizes from the history of past gradients and are the standard choice when only the gradient is available. A differentiable simulator provides more: the local curvature, contained in the Hessian $H_{ik}=\partial^2\mathcal{L}/\partial\theta_i\partial\theta_k$, is exactly the information needed to choose the step. Minimizing the local quadratic model of $\mathcal{L}$ gives Newton's step, $\bm\theta \leftarrow \bm\theta - H^{-1}\bm g$. Here $H^{-1}$ sets the step length separately in every direction, long where the loss is shallow and short where it is steep. Parameters of very different sensitivity are thereby placed on an equal footing, without a per-parameter learning rate to tune. Being local, the method refines the nearest minimum; the coarse grid and cone scans of Appendix~\ref{app:track_initial_guesses} supply the starting point for tracking.

Using the full Hessian directly has two drawbacks: its second derivatives are expensive, and far from the minimum it need not be positive-definite, in which case Newton's step could point away from the minimum rather than toward it. For a likelihood fit a well-known approximation removes both. Collecting the first derivatives of the per-sensor predictions into the Jacobian $J_{ji}=\partial\mu_j/\partial\theta_i$, the Hessian of Eq.~\ref{eq:counts_loss} separates into a term built purely from these first derivatives and a second term proportional to the residuals $(\mu_j-q_j)$ times second derivatives. Replacing the observed counts by their model means ($q_j\!\to\!\mu_j$, the \emph{expected} curvature) removes the residual term exactly and leaves the first weighted by the predicted mean rather than the observed count. What remains is the \emph{Gauss--Newton} matrix, which for the Poisson likelihood is
\begin{equation}
  F_{ik} \;=\; \sum_j \frac{1}{\mu_j}\,
       \frac{\partial\mu_j}{\partial\theta_i}\frac{\partial\mu_j}{\partial\theta_k}
   \;=\; 4\sum_j \frac{\partial\sqrt{\mu_j}}{\partial\theta_i}
                      \frac{\partial\sqrt{\mu_j}}{\partial\theta_k},
  \label{eq:fisher}
\end{equation}
the second form using $\partial\sqrt{\mu}/\partial\theta = (\partial\mu/\partial\theta)/ 2\sqrt{\mu}$. Working with $\sqrt{\mu}$ is the variance-stabilizing transformation for Poisson data, so that the sensor-to-sensor fluctuations no longer grow with the predicted charge. The full derivation, its identification with the Fisher information, and the family of least-squares losses that share this metric are given in Appendix~\ref{app:optimizer}. This matrix uses only first derivatives and is positive-semidefinite by construction, so the step is never an ascent direction. Being the expected rather than the observed curvature, it does not inherit the fluctuations of the residuals. For the Poisson likelihood it is moreover exactly the \emph{Fisher information matrix} of the model, and it measures how strongly the data constrain each parameter and each combination of parameters. The same construction applies to the other objectives of Sec.~\ref{sec:losses}, each contributing the Gauss--Newton matrix of its own residual, and for tracking the time term adds a second block so that $F$ is assembled from the derivatives of both the predicted charge and the predicted first-arrival model. Appendix~\ref{app:optimizer} gives these forms and their relation to Eq.~\ref{eq:fisher}.

\begin{figure}[htbp]
\centering
\includegraphics[width=0.49\textwidth]{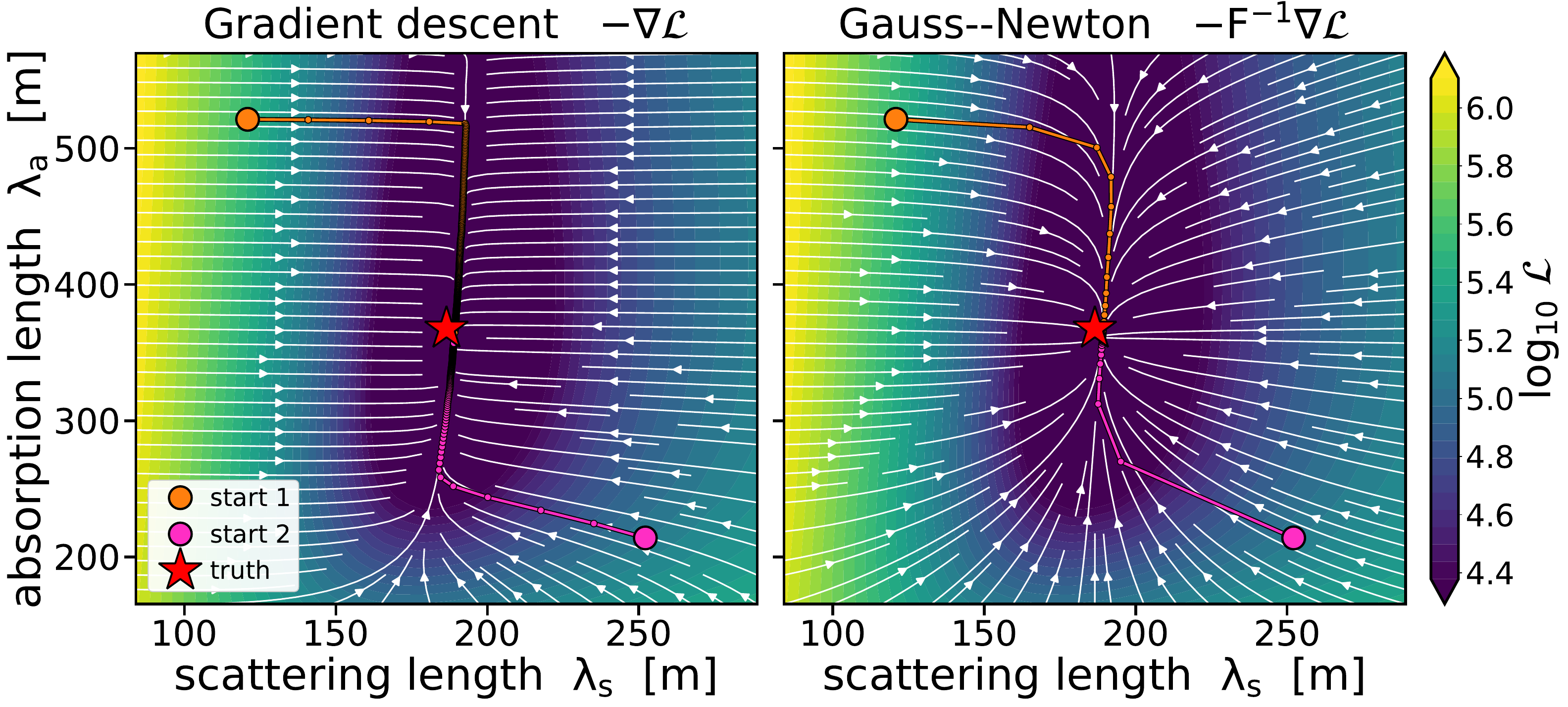}
\caption{The calibration loss over the scattering length $\lambda_s$ and the absorption length $\lambda_a$ at $405$~nm, near their true values (red star), for a single laser source. Compensating changes in the two leave the total collected light almost unchanged and differ only in how that light is distributed across the sensors, which is what opens the narrow valley. White streamlines show the normalized descent field. \textbf{Left:} raw gradient descent runs across the valley and along it, the signature of ill-conditioning. \textbf{Right:} the Fisher-preconditioned step rescales the two directions so that the streamlines point toward the minimum.}
\label{fig:loss_geometry}
\end{figure}

Replacing $H$ by $F$ gives the Gauss--Newton update $\bm\theta\leftarrow\bm\theta-F^{-1}\bm g$ (the Fisher-scoring, or natural-gradient, step). All parameters are updated simultaneously, and no per-parameter learning rate enters. Figure~\ref{fig:loss_geometry} contrasts raw gradient descent with this preconditioned step on the scattering and absorption lengths. The two are difficult to separate from the amount of collected light, because a change in one can be offset by a change in the other while the total charge stays nearly fixed. The loss therefore develops a long, narrow valley along that direction. What breaks the degeneracy is where the light lands rather than how much of it there is, since scattering redistributes charge across the detector while absorption removes it. Parameter pairs of this kind are common in the optical model, and the two shown here are an illustration rather than a special case; the preconditioned step treats them all alike.

The updates above require the Jacobian $J=\partial\mu/\partial\bm\theta$. In a stochastic simulator this would ordinarily be estimated by finite differences, re-running the simulation with each parameter perturbed. That estimate both amplifies the Monte Carlo noise and is biased at finite step size. It is the numerical differentiation a quasi-Newton fitter such as \textsc{minuit}~\cite{James:1975dr} performs. Instead we obtain $J$ by automatic differentiation, exactly and at a cost comparable to the simulation itself (Sec.~\ref{sec:performance}). In forward mode a single differentiated evaluation propagates the derivative for one parameter direction; the full Jacobian batches these into one vectorized pass, its cost a small multiple of one simulation because the fitted parameters are few. This is possible because the discrete scattering and surface choices are made differentiable by the score-function construction of Sec.~\ref{sec:photon_prop}.

Two refinements make the step reliable, and both fits use them identically. Far from the minimum the quadratic model is imperfect and the raw step can overshoot, so we damp it in the Levenberg--Marquardt~\cite{levenberg1944method, marquardt1963algorithm} manner, adding to $F$ a term proportional to its own diagonal and a smaller isotropic term. The first bounds the update along directions the data barely constrain while treating stiff and soft directions alike; the second guarantees the linear system is well posed. Because the parameters carry very different units and magnitudes, they are also rescaled to a common footing before the solve, either by fitting in log space or by dividing through by fixed characteristic scales. The values used, and the step-size control specific to each fit, are given in Appendix~\ref{app:optimizer}.

Each sensor carries a per-sensor efficiency $k_s$ multiplying the overall quantum efficiency $\mathrm{QE}(\lambda)$ of Eq.~\ref{eq:hit_charge}, so these parameters outnumber the bulk optical properties by orders of magnitude. They need not be fitted. Each one scales the prediction at its own sensor and nowhere else, so for given optical parameters the charge collected there fixes it in closed form. We use this to write the efficiencies as a function of the optical parameters and substitute them into the model, leaving the optical parameters as the only unknowns the optimizer carries. The Jacobian and the linear system are built over those alone, while the efficiencies follow from the same expression at every iteration and are reported at convergence. A gauge condition, $\langle\log k_s\rangle = 0$, applied by rescaling the solved values at each iteration, fixes the single overall normalization they share with the overall quantum efficiency and the source brightness, removing that exact degeneracy. The closed form and the numerical settings of the fit are collected in Appendix~\ref{app:optimizer}.

The matrix $F$ also gives access to the parameter uncertainties. Under the usual asymptotic assumptions and for a correctly specified model, its inverse $F^{-1}$ approximates the covariance of the estimate, the Cram\'er--Rao bound~\cite{cramer1999mathematical, rao1945information}. Because the Jacobian is already computed during the optimization, this estimate requires no additional simulation. We treat it as indicative here; a careful characterization of the parameter uncertainties, including residual model error, is left for future work.

\section{Calibration}
\label{sec:calibration}

Calibration aligns the tunable detector parameters $\bm\theta_{\mathrm{detector}}$ with control data, such that the simulation reproduces the response of the real detector. In our differentiable framework this is a gradient-based optimization with the Fisher--Gauss--Newton optimizer of Sec.~\ref{sec:gauss_newton}. The optical properties and efficiencies considered here are constrained by the charge alone, so the calibration fits use that observable and omit the timing term. The Fisher metric supplies the second-order step, which handles the strong correlations among the optical parameters.

The calibration combines two kinds of source. A collimated laser probes the medium along a well-defined path. Distributed isotropic sources illuminate the sensors over a range of positions and angles. Because the medium properties are wavelength-dependent (Sec.~\ref{sec:wavelength}), each source is used at five wavelengths, $337$, $375$, $405$, $445$ and $473$~nm, matching the dye-laser lines of the Super-Kamiokande water calibration~\cite{Abe:2013gga}.The optical constants take a separate value at each wavelength, while the reflectivities are achromatic. The per-sensor efficiencies are common to all of them. All are recovered in a single joint fit. Appendix~\ref{app:optimizer} gives the source positions, wavelengths, and photon budgets.

\subsection{Methodology}
\label{sec:calibration_method}

The studies below are designed as closure tests. The reference data are generated by the same simulator that is then fitted, which isolates the behavior of the estimator from any mismatch between the optical model and a real detector. They establish that the optimizer recovers the injected parameters and quantify the statistical precision, not the fidelity of the optical model to a real detector, which is left for validation against experimental data.

The fits minimize the Neyman $\chi^2$ of Eq.~\ref{eq:chi2_loss}, which for the calibration reads
\begin{equation}
\label{eq:calib_loss}
\mathcal{L}(\bm\phi) = \sum_{c}\sum_{s}
  \frac{\big[\,k_s(\bm\phi)\,m_{sc}(\bm\phi) - q_{sc}\,\big]^{2}}{q_{sc}},
\end{equation}
where $\bm\phi$ are the optical parameters, $c$ runs over the source and wavelength configurations and $s$ over the sensors, $m_{sc}$ is the predicted charge at sensor $s$ from configuration $c$ at unit efficiency, $q_{sc}$ the corresponding reference charge, and $k_s$ the efficiency of that sensor. Each residual is scaled by the square root of the reference charge, the nominal Poisson uncertainty on that measurement, so the weighting is fixed by the data rather than by the model.

The efficiencies $k_s$ are not free parameters. Each one scales the prediction at its own sensor and nowhere else, so it follows in closed form from the charge collected there (Sec.~\ref{sec:gauss_newton}), and the optimizer carries only $\bm\phi$.

The forward model is itself a Monte Carlo estimate, so the fit is built to average over that noise rather than to follow it. The step is assembled from two pieces, the residual $\bm r$ and its Jacobian $J$, combined as $\bm g = J^{\mathsf T}\bm r$ and $F = J^{\mathsf T}J$. These are evaluated on independent sets of simulated rays: the residual on a fresh realization at every iteration, so that the fit follows the expected response rather than the minimum of any one of them, and the Jacobian on a separate group of realizations. The gradient is a product of two estimated quantities, and estimating both from the same rays would bias it even though each is individually unbiased; drawing them independently removes that coupling~\cite{Nicolet2023Recursive}. The reported estimate is likewise an average over the tail of the trajectory rather than a single iterate. Appendix~\ref{app:optimizer} sets these out and gives what each is worth.

\subsection{Joint Optical and Efficiency Calibration}

We extract the bulk optical properties of the medium: the Rayleigh scattering length $\lambda_s$, the absorption length $\lambda_a$, and the wall and sensor reflectivities. These are strongly correlated, since increasing either scattering or absorption attenuates the collected light, but they leave distinct spatial imprints on the charge pattern, which the fit exploits through the off-diagonal structure of the Fisher metric. A central laser positioned on the top end-cap and directed along the axis provides a controlled, anisotropic illumination in which the direct beam constrains the absorption while the off-axis scattered and reflected light constrains the scattering length and the reflectivities. The reflection of each surface is modeled as a mixture of a specular and a diffuse (Lambertian) component (Sec.~\ref{sec:photon_prop}), parametrized by a reflectivity and a specular fraction for the walls and for the sensors, both recovered from the charge pattern.

\begin{figure*}[htbp]
\centering
\includegraphics[width=\textwidth]{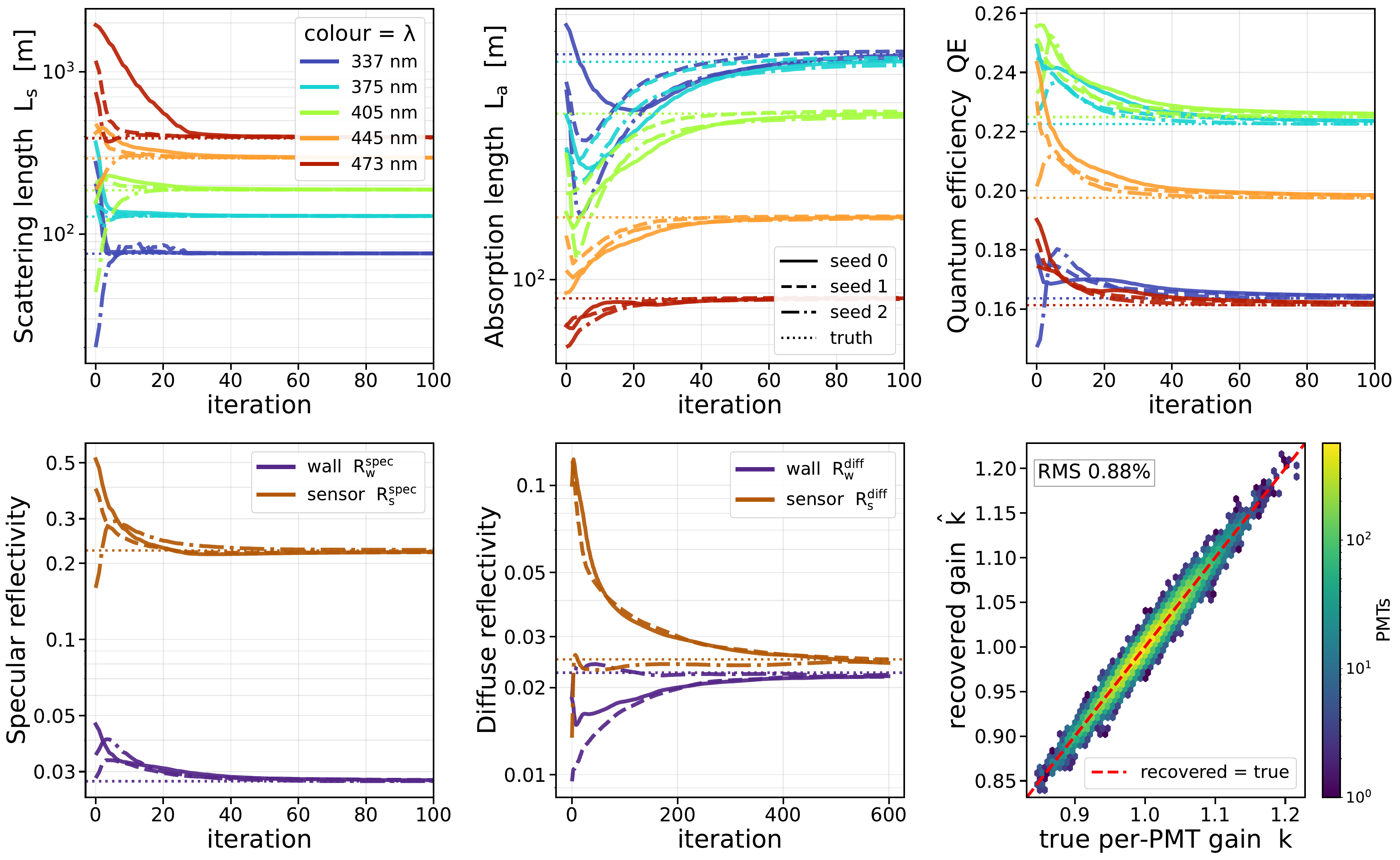}
\caption{Full calibration convergence. A single joint Gauss--Newton fit recovers the complete optical model, run from three perturbed starting points (line style; dotted lines mark the truth). \textbf{Top:} the per-wavelength scattering length $\lambda_s$, absorption length $\lambda_a$, and quantum efficiency at the five calibration wavelengths (color). \textbf{Bottom:} the wavelength-independent specular and diffuse reflectivities of the walls and sensors, and the recovery of the full set of per-sensor efficiencies (recovered versus true, RMS $0.9\%$). Curves are shown as a running mean over the trajectory, and the horizontal axis is truncated at $100$ iterations for the parameters that settle early while spanning the full $600$ for the diffuse reflectivities, which are the slowest. The optical constants converge to the underlying Super-Kamiokande water model and the reflectivities to their injected values.}
\label{fig:calib_wavelength}
\end{figure*}

Because the optical constants are wavelength-dependent, the effective lengths recovered are those at each source wavelength, and a scan over several laser wavelengths traces out the wavelength dependence of the medium. Starting from a range of initial values, the fit converges to the true optical parameters at each wavelength, disentangling the correlated attenuation effects by using the full spatial charge pattern rather than the total collected light alone. Figure~\ref{fig:calib_wavelength} shows one such joint fit converging from perturbed starting values to the true per-wavelength scattering length, absorption length, and quantum efficiency, together with the shared wall and sensor reflectivities, their specular and diffuse components, and the full set of per-sensor efficiencies. The recovered wavelength dependence reproduces the underlying Super-Kamiokande water model. This is directly analogous to the dye-laser calibration used in operating water-Cherenkov detectors.

The same fit determines the efficiency of every individual photosensor, of order $10^4$ parameters in the SK-like geometry. The isotropic sources are what make this possible. Their overlapping coverage reaches each sensor from several directions, which separates a genuine efficiency difference from the attenuation along any one path. Each efficiency scales the prediction at its own sensor alone, so the closed form of Sec.~\ref{sec:gauss_newton} supplies the full set of them at every iteration, and the recovery of so many parameters stays tractable. The gauge condition holds them to unit geometric mean, so the absolute scale is carried by the overall quantum efficiency and only relative differences between sensors are attributed to the per-sensor factors. The recovered values track the injected ones with a residual spread of $0.9\%$, set by the available photon statistics.

\section{Tracking}
\label{sec:tracking}

This section applies the same differentiable machinery to the inverse problem of inferring a particle's trajectory from the light it produces.

\begin{figure*}[htbp]
\centering
\caption{Two-dimensional loss landscape for different parameter combinations, with associated gradient streamlines, as observed in a randomly selected data-like event. The true parameter values are indicated by red stars.}
\label{fig:trk_grads}
\includegraphics[width=0.99\textwidth]{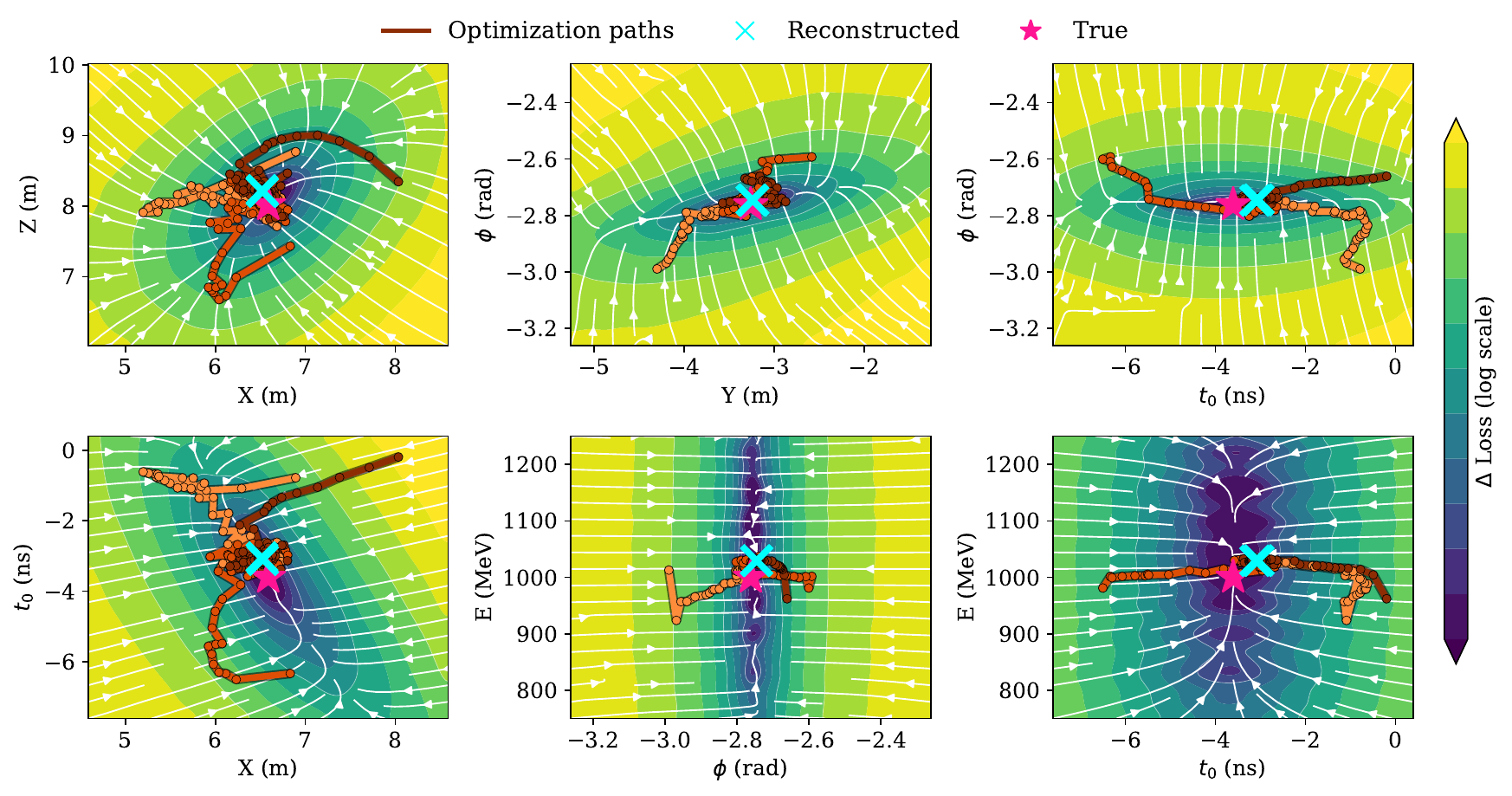}
\end{figure*}

\begin{figure}[h!]
\centering

\caption{Evolution of reconstruction errors for 1~GeV muons over optimization iterations for the track vertex position, direction, interaction time ($t_0$), and particle momentum. The histograms on the right of each panel show the distribution of the final reconstructed values.}
\label{fig:tracking_convergence_SK_like}
\includegraphics[width=0.49\textwidth]{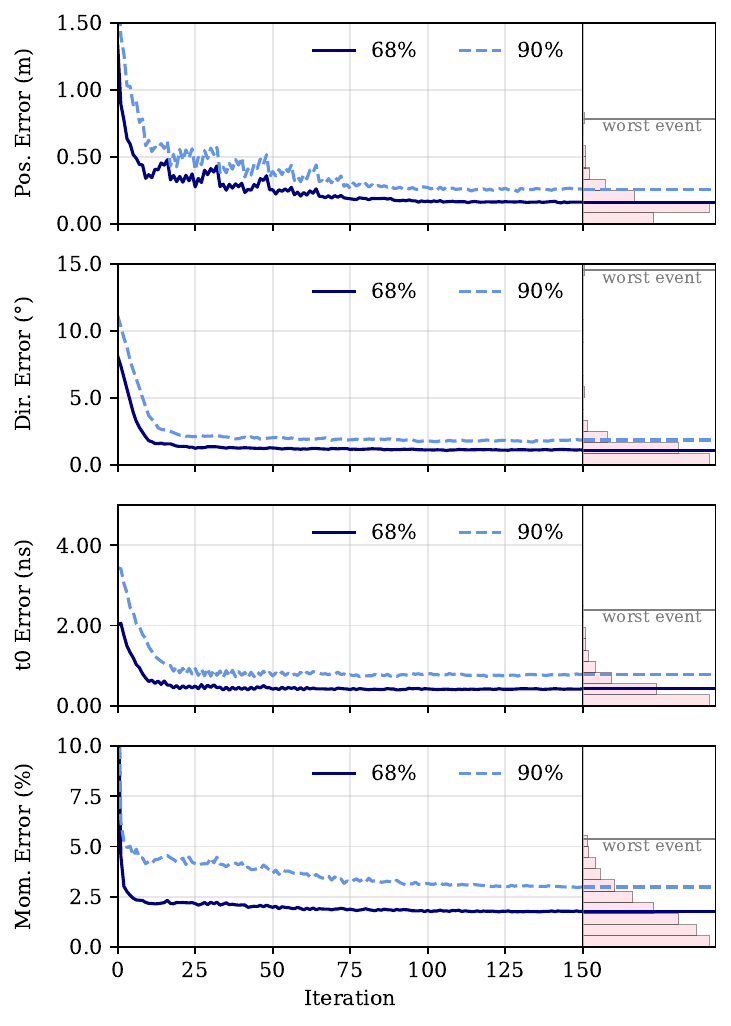}
\end{figure}

\begin{figure}[htbp]
\centering
\caption{Tracking performance as a function of the number of rays, $N_{\mathrm{rays}}$.}
\label{fig:trk_perf_vs_nrays}
\includegraphics[width=0.49\textwidth]{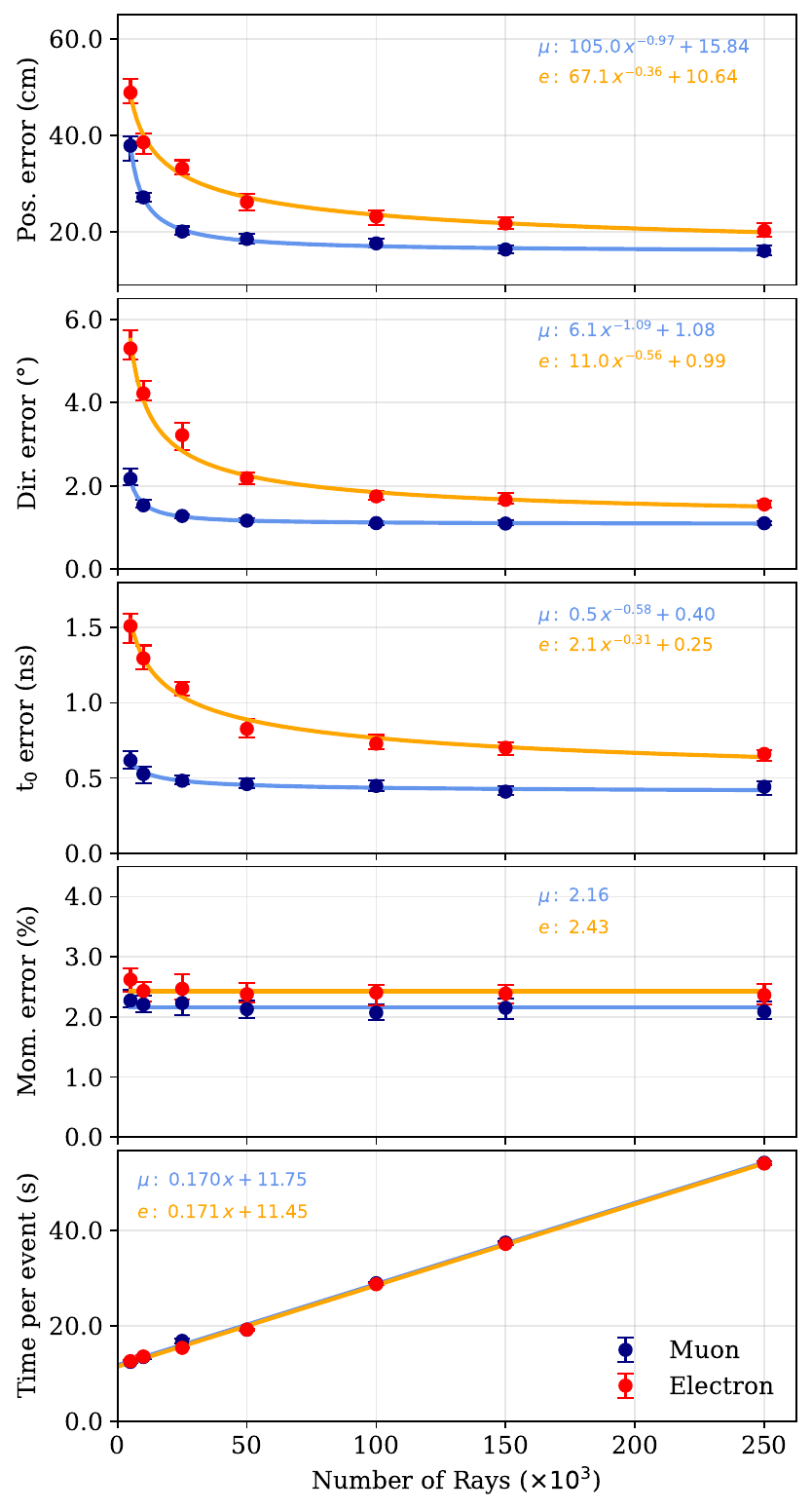}
\end{figure}

The differentiable simulator developed in the previous sections enables gradient-based reconstruction of particle tracks. Given a set of observed detector hits, we infer the track that best explains the data by minimizing the joint charge and time loss of Eq.~\ref{eq:total_loss} over the seven track parameters: the vertex position $\mathbf{x}$, the direction $(\theta,\phi)$, the start time $t_0$, and the energy $E$. The direction is carried in the fit as the unnormalized pairs $(\sin\theta,\cos\theta)$ and $(\sin\phi,\cos\phi)$ rather than as the angles themselves, so that the parameter vector has nine components for seven degrees of freedom. An explicit angle has a branch cut, where the gradient is discontinuous; the redundant encoding has none. The pairs are renormalized when the angles are read back, and the two flat directions this introduces are bounded by the isotropic damping term of Sec.~\ref{sec:gauss_newton}. As in the calibration of Sec.~\ref{sec:calibration}, this is a gradient-based optimization solved with the Fisher--Gauss--Newton optimizer of Sec.~\ref{sec:gauss_newton}, so we describe here only what is specific to tracking. Reconstruction uses both observables rather than the charge alone: the Fisher metric is assembled by automatic differentiation from the derivatives of the predicted charge and first-arrival time with respect to the track parameters, with the charge term fixing the energy scale and the timing term constraining the vertex, direction, and start time. Smooth gradients in the vertex and direction moreover require the photon-relaxation model of Sec.~\ref{sec:photon_relaxation}, which gives each ray a soft sensor footprint in place of the strict binary intersection retained for calibration. The damped second-order step then updates all seven parameters simultaneously, with no per-parameter learning rate, and the reported estimate is the Polyak average of the final iterates. All tracking studies use the wavelength-dependent water optical properties of Super-Kamiokande~\cite{Abe:2013gga}, with the wall and sensor reflection described by the specular and diffuse mixture of Sec.~\ref{sec:photon_prop} at total reflectivities of $5\%$ and $25\%$ and specular fractions of $0.55$ and $0.90$. These are the same optical parameters that serve as the calibration targets, so the detector model is consistent between the calibration and tracking studies.

Figure~\ref{fig:trk_grads} illustrates the optimization trajectories for a representative data-like event, projected onto several two-dimensional parameter slices. Each trajectory corresponds to a different random initialization of the track parameters. Regardless of the starting point, the optimizer consistently converges to the correct solution through gradient-based minimization. Convergence is typically achieved even for initial conditions far from the true values, except in cases where the predicted light reflection pattern closely resembles the true event topology. In such cases, the optimizer may settle in a local minimum. These ambiguities are easily mitigated by performing a coarse grid search over the initial parameter space, as detailed in Appendix~\ref{app:track_initial_guesses}.

For quantitative performance studies, we reconstruct 500 muon events with $E_\mu = 1000$~MeV (1~GeV) simulated in the SK-like detector. Events are given an isotropic direction and displaced uniformly within a cylinder of $12$~m in radius and half-height, with placements rejected and redrawn unless the projected end of the track falls within $95\%$ of the detector's linear dimensions.Both the data-like samples and the differentiable expectation are propagated with $K = 8$ iterations, sufficient for the residual light to be negligible (Appendix~\ref{app:Kconvergence}).

Figure~\ref{fig:tracking_convergence_SK_like} shows the convergence behavior of the position, direction, $t_0$, and momentum errors as a function of optimization iteration. Good convergence is reached for $N_{\mathrm{rays}} \sim 150$k, with fewer than 100 iterations sufficient for stable convergence. This corresponds to a total reconstruction time of approximately 30~seconds per event, comparable to state-of-the-art alternatives.

The reconstruction performance as a function of $N_{\mathrm{rays}}$ is summarized in Fig.~\ref{fig:trk_perf_vs_nrays}. For muons we obtain a vertex resolution of approximately $16.1^{+1.4}_{-0.9}$~cm, an angular resolution of $(1.10^{+0.05}_{-0.06})^{\circ}$, and a momentum resolution of $2.1^{+0.2}_{-0.1}$\%. For electrons we obtain $20.2^{+1.6}_{-1.2}$~cm, $(1.56^{+0.09}_{-0.08})^{\circ}$, and $2.4\pm0.2$\%. These results are comparable to those achieved by FitQun (15.8~cm, 1.0$^\circ$, 2.3\% for muons; 20.6~cm, 1.5$^\circ$, 2.9\% for electrons), when evaluated on a similar simulated data set~\cite{Super-Kamiokande:2019gzr}. While this comparison should be interpreted qualitatively, since our detector geometry, sensor response, and medium properties do not exactly replicate those of Super-Kamiokande, the close agreement demonstrates that the proposed differentiable approach can reach performance levels comparable to state-of-the-art, non-differentiable methods. At the same time, it offers improved computational efficiency, calibration capabilities and the additional benefits discussed in the introduction. Moreover, the error distributions remain consistently narrow across all tested events, with no observed reconstruction failures and only small tails. These tails are comparable to those observed in FitQun~\cite{Jiang:2019vpo} and are consistent with unlikely data-like events, such as large early scattering of the muon that are not well represented by the average prediction. This indicates stable optimization behavior and suggests that the approximations introduced in LUCiD do not lead to any appreciable degradation of the reconstruction performance when compared to FitQun.

\begin{figure}[h!]
\centering
\caption{Tracking performance as a function of the simulated muon energy.}
\label{fig:trk_perf_vs_energy}
\includegraphics[width=0.49\textwidth]{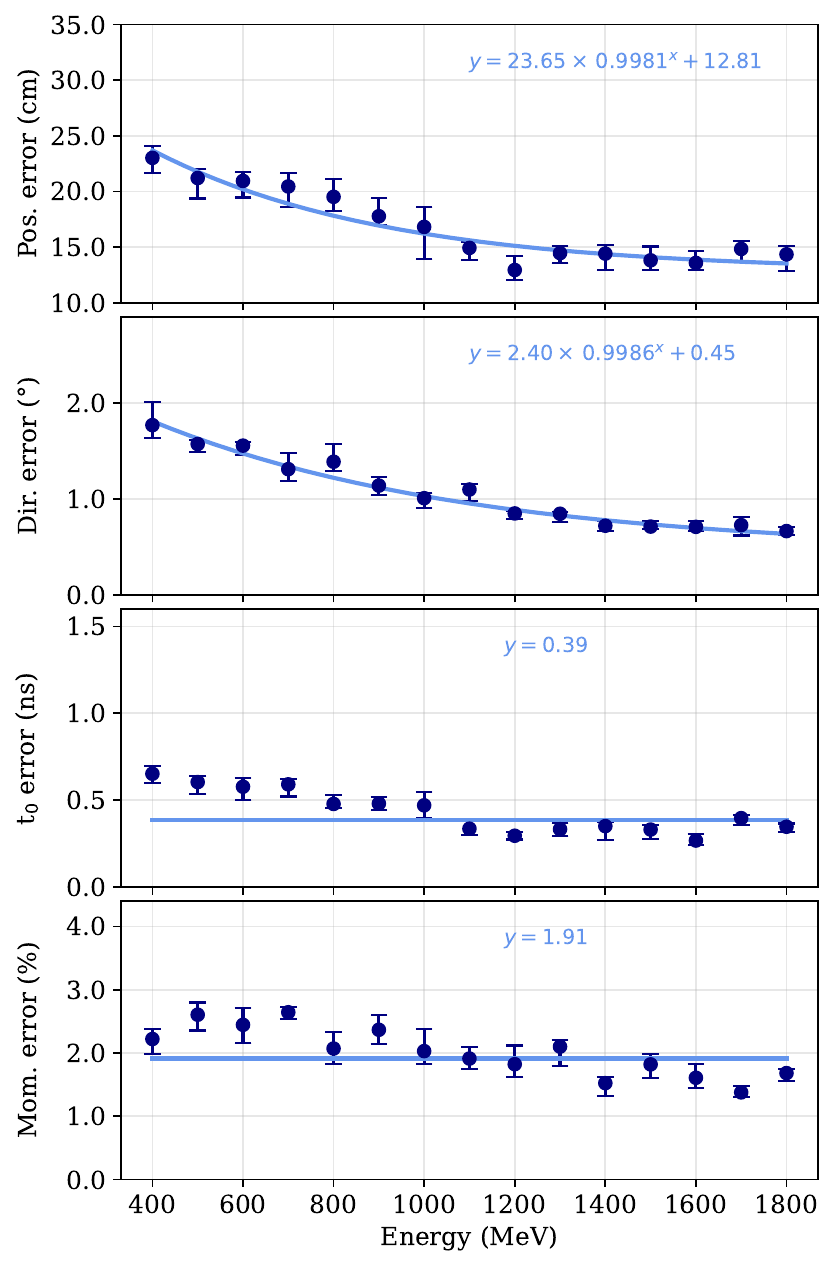}
\end{figure}

The performances of the track reconstruction vary smoothly with the simulated track energy, see Figure~\ref{fig:trk_perf_vs_energy}, and demonstrate that the simulator already performs well across a wide range of energies.

\section{Application to Detector Design}
\label{sec:det_design}

\begin{figure}[h!]
\centering
\caption{Tracking performance as a function of the number of photosensors.}
\label{fig:trk_perf_vs_sensors}
\includegraphics[width=0.49\textwidth]{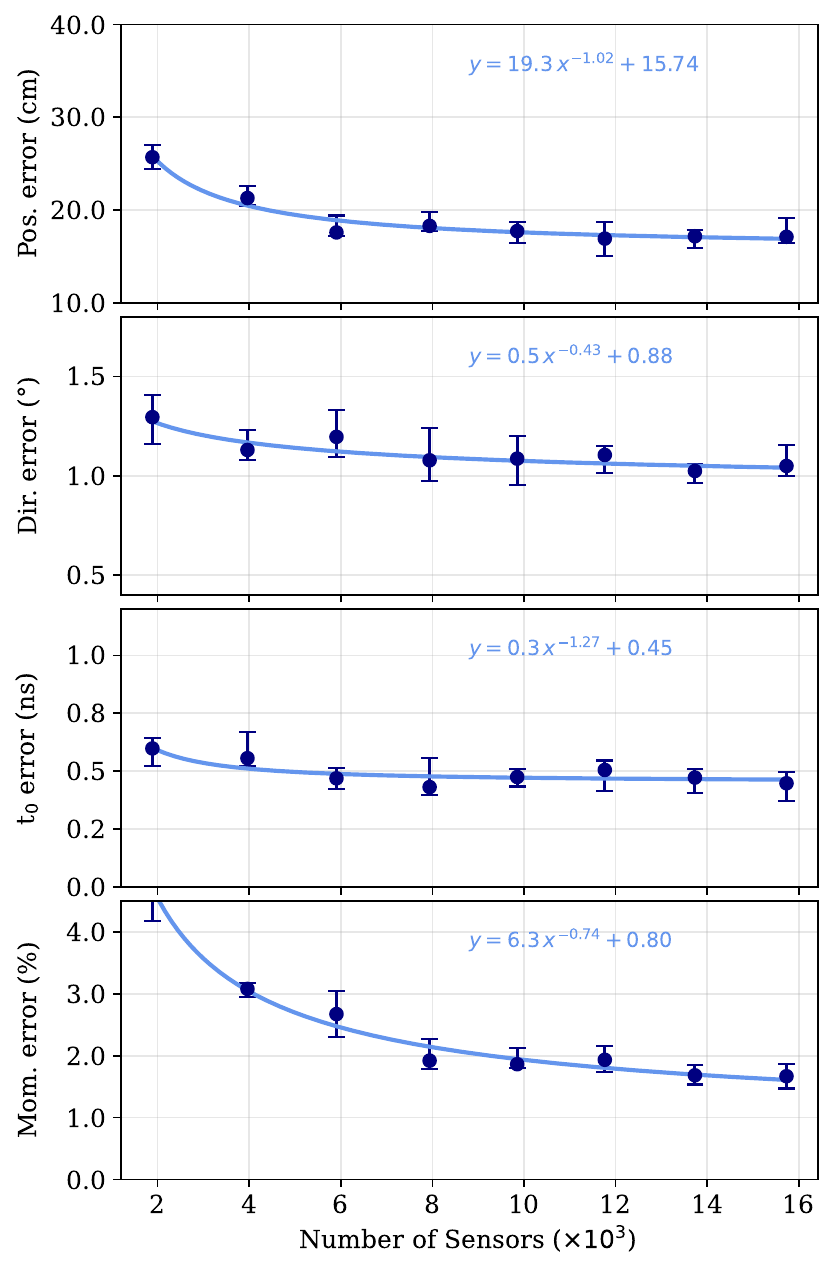}
\end{figure}

A persistent challenge in detector design optimization is that the key performance metrics are typically reconstruction metrics, such as those presented in the previous section, yet reconstruction algorithms are often tuned to a given detector geometry. Adapting them to alternative designs is therefore a labor- and compute-intensive task. For example, FitQun avoids explicit ray tracing by relying on parametrizations derived from sensor-by-sensor lookup tables. A limitation of this approach is that these tables and parametrizations are valid only for a fixed geometry, requiring the entire reconstruction procedure to be re-tuned and re-validated whenever the detector design changes. In contrast, our differentiable framework performs ray tracing on the fly for arbitrary detector configurations, enabling direct comparison of reconstruction performance across different geometries without any re-training or re-tuning. This capability makes detector design optimization straightforward: performance metrics can be evaluated consistently for a standard dataset while varying geometry or sensor parameters. Figure~\ref{fig:trk_perf_vs_sensors} illustrates this concept using the same configuration as in Fig.~\ref{fig:trk_perf_vs_nrays}, but varying the number of photosensors in the SK-like detector. The same strategy can be generalized to explore multi-dimensional design spaces, for instance by varying sensor detection efficiency, timing and photoelectron resolution, detector size, or shape. This provides a practical pathway to detector design within experimental and budgetary constraints.

\section{Future directions}
\label{sec:future_dir}
\subsection{Complex Topologies}
\label{sec:future_dir_topos}

Although multi-particle optimization is left for future work, the losses described above extend naturally to events involving multiple simultaneous particles, requiring no conceptual modification of the framework: the expected charges and times for the full event are differentiable combinations of the corresponding single-particle quantities, so in practice one simply evaluates the hit-making logic for the rays of all particles rather than for a single one. The gradients obtained through backpropagation then point in directions that simultaneously improve the reconstructed parameters of all particles, automatically accounting for overlapping light patterns. The initialization of Appendix~\ref{app:track_initial_guesses} would need to be generalized, since the fit refines the nearest minimum. The vertex and $t_0$ search relies on arrival times and, for particles sharing a common interaction vertex, would be performed once per event; the directional scan would instead seek sequentially the several best local minima rather than only the global one, each seeding one particle hypothesis.

A related extension concerns tracks with hard scatters, particularly relevant for charged pions undergoing hadronic interactions. Such events can be modeled as multiple connected track segments sharing a common vertex at the scattering point, with gradients naturally favoring kink locations that best explain the observed light distribution. Reconstructing such topologies also raises the question of model selection: the simplest approach is to perform reconstruction under several particle-number hypotheses and compare results using likelihood ratios or information criteria. However, gradient information may also inform model selection, by initializing with excess particles whose contributions are suppressed during optimization if not supported by the data.

\subsection{Extension to Complex Geometries}

The geometries supported in this work are described by parametric templates: the closed cylinder, sphere, and box boundaries, treated with analytic ray--surface intersection, and the volumetric string arrays, treated with per-module closest-approach tests. Detectors whose boundaries do not reduce to these templates, such as irregular vessels, internal reflectors, or support structures, motivate a more general surface description.

For detectors with multiple instrumented regions, such as an Inner Detector (ID) surrounded by an Outer Detector (OD) or external cosmic-ray veto systems, modest extensions to the present framework would be sufficient. These could involve either expanding the differentiable geometry to include multiple coupled volumes with distinct optical and sensor properties, or combining the reconstruction of the primary target volume with auxiliary veto information at a separate stage.

To accommodate arbitrarily complex detector geometries more generally, a promising approach is to represent surfaces as neural implicit functions, following recent advances in neural radiance fields~\cite{Mildenhall20nerf}. Such representations can learn complex shapes from CAD models or calibration data while remaining fully differentiable, enabling gradient-based optimization without requiring analytic intersection formulas for each component. This would also allow modeling of realistic, non-spherical PMT geometries, which affect both acceptance and timing response.

This flexibility comes with a tradeoff: detailed geometric representations require additional intersection evaluations, increasing computational cost. Balancing geometric fidelity against reconstruction speed will be an important consideration for applying this framework to detectors with complex optical module designs, multi-region instrumentation, or irregular layouts.

\subsection{Extended Detector Modeling}

The current implementation treats photosensors as ideal spherical surfaces and produces hit-level observables. The timing observable is then the first arrival, a single order statistic per sensor, which fluctuates more than the collected charge and makes the timing term correspondingly harder to optimize. Extending to waveform-level predictions requires addressing the discontinuity of time binning, which can be achieved through kernel density estimation to maintain differentiability. This would enable more sophisticated timing estimators that extract information directly from waveform features.

Real PMTs exhibit complex quantum efficiency and photoelectron transit time profiles that depend on the photon incident position and direction. The modular architecture of the framework naturally accommodates learning the PMT response maps as a function of multiple variables from laboratory measurements or detailed simulations using a network architecture similar to the SIREN employed for Cherenkov emission (Sec.~\ref{sec:ray_generation}). Once trained, this component integrates directly after the photon ray tracing in the PMT hit making step. For modules housing multiple PMTs, as those to be used in Hyper-Kamiokande and IceCube, the same approach extends to learning a response map that distributes an incident photon's contribution across the constituent sensors based on where it strikes the outer enclosure.

This modularity enables a practical two-stage calibration strategy. First, a base PMT response map is trained from laboratory characterization of a reference sensor, capturing the intrinsic response shared by all units of the same type. Then, during in-situ calibration, the full PMT response map for each installed sensor can be refined using calibration sources rather than being limited to simple scaling as in conventional approaches. The differentiable framework makes this higher-dimensional optimization tractable.

The same reasoning applies to the calibration sources themselves. Their intensity, angular emission profile, and pointing enter the simulation as differentiable inputs, so they can be profiled as nuisance parameters alongside the detector optics and sensor response, relaxing the need for precise a~priori characterization of the sources. A full joint in-situ fit of the source and detector properties is a natural direction for future work.

Real detector media exhibit spatial variations in optical properties due to impurities, temperature gradients, or material aging effects. These inhomogeneities affect light propagation and visibility but are difficult to model beyond simple parameterizations. A SIREN-like network can represent spatially and temporally varying scattering and absorption lengths as continuous fields, with rays sampled volumetrically to compute path-dependent attenuation. Such a model could be learned directly from calibration data, capturing detector inhomogeneities and their evolution over time.

\subsection{Stochastic Extensions}
\label{sec:future_dir_stochasticity}
The simulation operates in expected-value mode throughout, propagating intensity-weighted rays rather than sampling individual photon fates. However, certain processes exhibit irreducible stochasticity that may be important to capture. Examples include the distribution of outcomes when photons undergo multiple internal reflections within a photosensor assembly, and particle-level fluctuations, such as electromagnetic showers and inelastic interactions, in light emission.

Generative models such as normalizing flows~\cite{PapamakariosNormalizingFlows} provide a natural framework for differentiable stochastic simulation. These approaches structure the random sampling process in a way that allows gradients to propagate through, enabling end-to-end training against detector data. This would allow learning physical fluctuation distributions directly from observation rather than relying solely on first-principles simulation.

\subsection{Towards End-to-End Differentiable Analysis}

Beyond unifying traditionally fragmented workflows, the differentiable framework enables capabilities that were previously impractical. Automatic differentiation provides not only gradients but also efficient Hessian computation, giving access to curvature information for uncertainty estimation. This allows calibration uncertainties to propagate through reconstruction into physics results while preserving correlations, whereas current approaches typically lose information at each stage boundary.

The differentiable detector simulation presented here forms one component of a potentially fully differentiable analysis chain. Upstream, this could connect to differentiable event generation and interaction physics, propagating gradients from theory parameters through particle production into detector response. Downstream, extension to differentiable likelihood construction and sensitivity metrics would enable optimization from detector configuration through to physics results. Such a complete pipeline would allow gradients to flow from fundamental physics parameters to final measurements.

Finally, the modular architecture facilitates extension to detectors with multiple readout modalities. Liquid argon time projection chambers, for example, provide both scintillation light and ionization charge. Differentiable simulations developed independently for each modality can be composed within a unified framework, enabling joint reconstruction and co-optimization of how information from each readout is combined.\\

\section{Conclusions}
In conclusion, we have introduced \lucid, the first end-to-end differentiable simulation framework for large-scale homogeneous optical particle detectors that unifies light generation, propagation, and sensor response within a single gradient-based optimization pipeline. By targeting smooth, differentiable expectations in detector-output space while retaining high-fidelity validation through a complementary data-like Monte Carlo mode, \lucid enables calibration and reconstruction to be performed simultaneously and efficiently. This directly challenges the long-standing state-of-the-art approach in experimental particle physics, which relies on non-differentiable sampling-based simulators for simulation, calibration and reconstruction procedures.

We demonstrated that physically meaningful and stable gradients can be obtained across all key stages of optical detector response, including light emission, transport, medium interactions, boundary reflections, and sensor acceptance. These capabilities enable robust gradient-based extraction of correlated medium parameters, high-dimensional per-sensor efficiency calibration at the scale of ten thousand parameters, and competitive track reconstruction performance in an SK-like detector geometry. In all cases, the differentiable approach matches or surpasses conventional methods in accuracy while offering a simpler, more unified, and computationally efficient workflow.

Taken together, these results constitute the first demonstration that end-to-end differentiable simulation is not only viable but practically ready for use in ongoing and future optical detector experiments. By removing artificial boundaries between simulation, calibration, and reconstruction, this work provides the essential ingredients for a paradigm shift in how detector modeling and data analysis are performed in particle physics, opening the door to fully integrated optimization, uncertainty propagation, and detector design within a single coherent framework.

\section*{Acknowledgments}
We acknowledge valuable discussions within the CIDeR-ML collaboration, especially with K.M. Tsui, M. Wilking, and J. Xia, on conventional simulation and water Cherenkov detector modeling. This work was supported by the Center for Data-Driven Discovery, Kavli IPMU (WPI), JSPS KAKENHI Grant Number JP24K23938, the Ozaki Exchange Program, the U.S. Department of Energy, Office of Science, Office of High Energy Physics, and U.S.-Japan Science and Technology Cooperation Program under Contract DE-AC02-76SF00515.

\hfill

\bibliographystyle{apsrev4-1}
\bibliography{biblio}

\pagebreak

\newpage
\appendix
\section{SIREN Training Details}
\label{app:siren_training}

The surrogate model $f^{\mathrm{Cherenkov}}_\gamma(E, s, \theta)$ is implemented as a Sinusoidal Representation Network (SIREN)~\cite{sitzmann2020implicit}, chosen for its ability to represent high-frequency features smoothly and differentially.

We generate the training data using \textsc{PhotonSim}, a GEANT4-based tool that simulates photon emission for monoenergetic charged particles in water. For each particle type, photon distributions are computed on a non-uniform energy grid of 322 (339) energy points for muon (electrons) spanning from the lowest simulated energies up to 100~GeV (10~MeV spacing at the lowest energies, widening to $\sim$10~GeV near the upper end). Each energy point comprises $10^4$ simulated events, from which we extract a $500 \times 500$ two-dimensional histogram in the photon emission angle $\cos\theta$ and the normalized emission distance $s/s_{\max}(E)$, where $s_{\max}(E)$ is the energy-dependent characteristic track length introduced in Sec.~\ref{sec:ray_generation}. The resulting 3D lookup table $(E,\, \cos\theta,\, s/s_{\max})$ is log-transformed and normalized to enhance sensitivity to low-intensity regions.

The SIREN architecture consists of three input features $(E, \cos\theta, s/s_{\max})$, an input sinusoidal layer followed by three hidden layers of 256 neurons each with sinusoidal activations of frequency $\omega_0 = 30$, and a single linear output neuron. The output is squared to ensure non-negative photon intensities. The inputs are min-max normalized to the $[-1, 1]$ range, while the targets are log-transformed as $\log_{10}(y + \delta)$ with $\delta = 10^{-3}$ and then min-max normalized to $[0, 1]$. The offset $\delta$ also sets the threshold below which a histogram bin is treated as empty: all bins above it are kept, together with a randomly selected half of the empty bins, retained as anchors so the surrogate learns the extent of the zero-intensity regions.

Training is performed in JAX using the Adam optimizer with an initial learning rate of $10^{-4}$ (reduced on plateau down to $5\times10^{-7}$ by a patience-based scheduler) and a global gradient-norm clip of 1.0. Training batches are drawn with each energy weighted equally across the grid, and within each energy 20\% of the sampling weight is allocated to the high-intensity bins (the Cherenkov ring) with the remaining 80\% uniform, balancing the otherwise low-occupancy regions. The batch size is 25{,}000 elements, and training is conducted for $1.5\times10^{5}$ steps, sufficient for the mean squared error (MSE) loss to reach a stable plateau. We use a 99/1 train-validation split and store the model parameters at the minimum validation loss. The same procedure and hyperparameters produce the $dE/dx$ surrogate used by the scintillation model of Appendix~\ref{app:scintillation}, with the $\cos\theta$ axis replaced by $dE/dx$ and each histogram entry weighted by the energy deposited in the corresponding step.

\section{A surrogate for particle light emission in scintillator media}
\label{app:scintillation}

The differentiable modeling of light emission in scintillator media is achieved analogously to the Cherenkov light case. First, a SIREN surrogate is trained to interpolate a lookup table over $(E,\, dE/dx,\, s/s_{\max}(E))$, generated with \textsc{PhotonSim}, in which each entry is weighted by the energy deposited in the corresponding \textsc{Geant4} step. The marginal of this surrogate over the $dE/dx$ axis is then the deposited-energy profile $E_{\rm dep}(s; E_{\rm track})$ along the track, while its energy-weighted mean over $dE/dx$ gives the local energy-loss rate at each $s$. In contrast to the Cherenkov case, the photon emission angle does not need to be modeled, because scintillation light is emitted isotropically; the second surrogate axis instead encodes the local $dE/dx$, which is required to evaluate the light yield in the presence of quenching. Rays are then distributed along the track with a density proportional to $E_{\rm dep}(s; E_{\rm track})$, so that they concentrate where the energy is deposited, and each of the $N$ rays carries an equal share $E_{\rm track}/N$ of the particle energy with an intensity given by the quenched light yield
\begin{equation}
I^{\mathrm{scint}}_i = \frac{S\,(E_{\rm track}/N)}
{1 + k_B\,(dE/dx)_i + C\,(dE/dx)^2_i},
\end{equation}
where $(dE/dx)_i$ is the local energy-loss rate at the ray position and the denominator implements Chou's law,
\begin{equation}
\frac{dL}{dx} = S\,\frac{dE/dx}{1 + k_B\,(dE/dx) + C\,(dE/dx)^2},
\end{equation}
a generalization of Birks' law in good agreement with data~\cite{Caravaca:2020lfs} in which $S$ is the light yield in the linear (unquenched) regime, $k_B$ is the Birks constant, and $C$ is a second-order quenching constant; setting $C = 0$ recovers Birks' law. We model the emission time of each scintillation ray combining the same arrival-time baseline used for Cherenkov light, $f^{t_0}_\gamma(E_{\rm track}, s)$ of Eq.~\ref{eq:t0_correction}, with a stochastic de-excitation delay drawn from a rise-and-fall (bi-exponential) profile
\begin{equation}
g(t;\tau_{\rm rise},\tau_{\rm fall}) =
\frac{e^{-t/\tau_{\rm fall}} - e^{-t/\tau_{\rm rise}}}
{\tau_{\rm fall} - \tau_{\rm rise}},
\end{equation}
characterized by a fast rise time $\tau_{\rm rise}$ and a slower decay time $\tau_{\rm fall}$, using the parametrization in Ref.~\cite{Caravaca:2020lfs}. The delay is sampled through the reparameterization $t = -\tau_{\rm rise}\ln U_1 - \tau_{\rm fall}\ln U_2$, with $U_1, U_2 \sim \mathcal{U}[0,1]$, so that it remains differentiable with respect to both time constants. Each ray is additionally assigned an isotropic emission direction and an emission wavelength sampled from the scintillator emission spectrum, modeled as a Moyal distribution truncated to the medium's $[\lambda_{\min}, \lambda_{\max}]$ window. Through this formulation, the expected detector response depends explicitly on the light yield $S$, the quenching constants $k_B$ and $C$, and the de-excitation times $\tau_{\rm rise}$ and $\tau_{\rm fall}$, allowing all of them to be tuned using gradient-based optimization. In a scintillating medium such as a water-based liquid scintillator (WbLS), these scintillation rays are generated alongside the Cherenkov rays and the two contributions are summed, following $f^{\rm Particle}_\gamma = f^{\mathrm{Cherenkov}}_\gamma + f^{\mathrm{Scintillation}}_\gamma$.

\begin{figure}[htbp]
\centering
\caption{Example of a 1~GeV muon data-like Cherenkov (red) vs scintillation (blue) light fraction compared to the matching differentiable prediction in a JUNO-sized detector using 5\% WbLS as detector medium.}
\label{fig:wbls_example}
\includegraphics[width=0.49\textwidth]{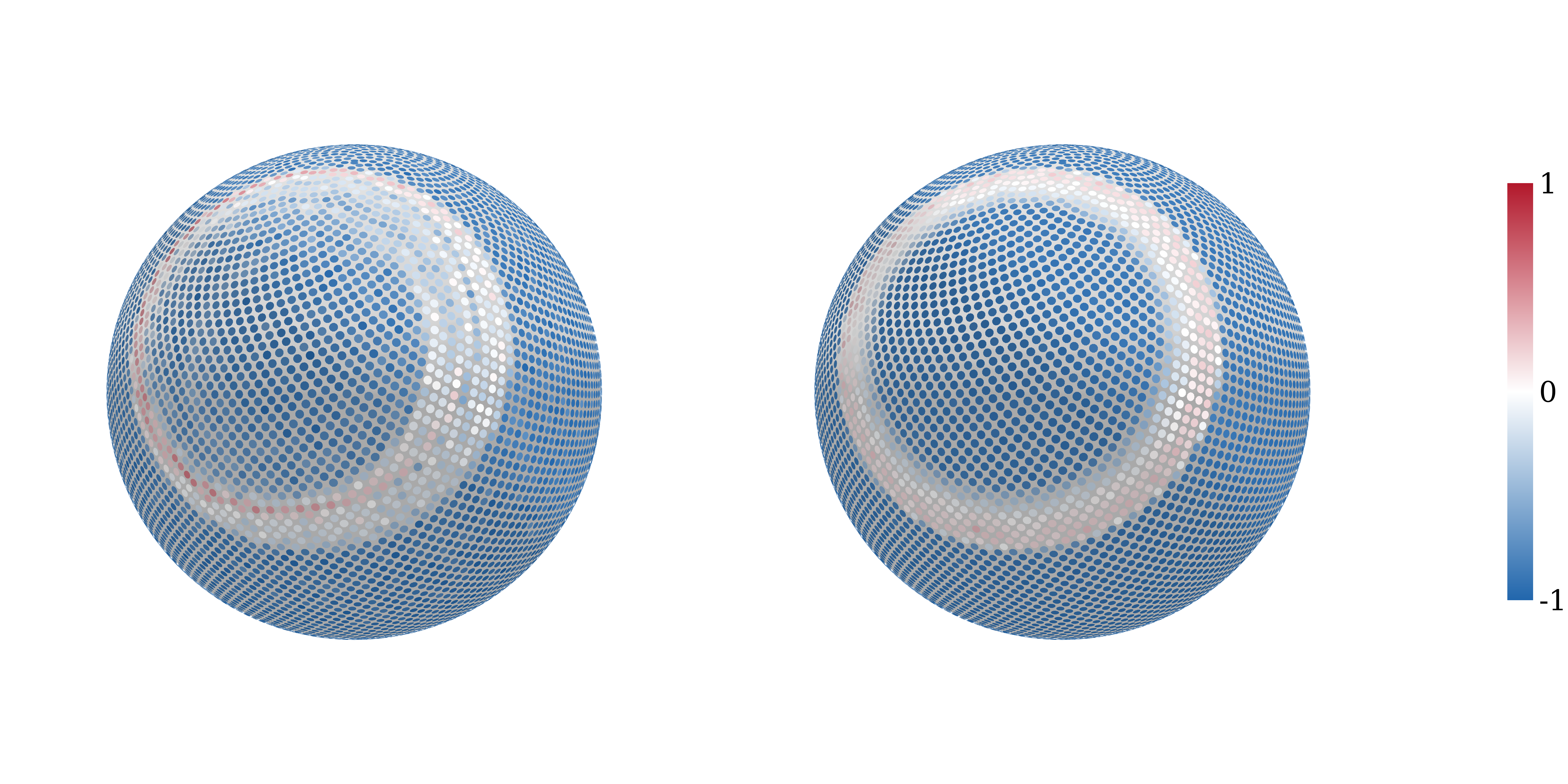}
\end{figure}

\begin{figure}[htbp]
\centering
\caption{Likelihood scan and associated gradients of the scintillation yield scale $S$ for the data-like event in Fig.~\ref{fig:wbls_example}. The loss is a negative-log likelihood in the observed vs predicted sensor photoelectrons.}
\label{fig:scint_diff}
\includegraphics[width=0.46\textwidth]{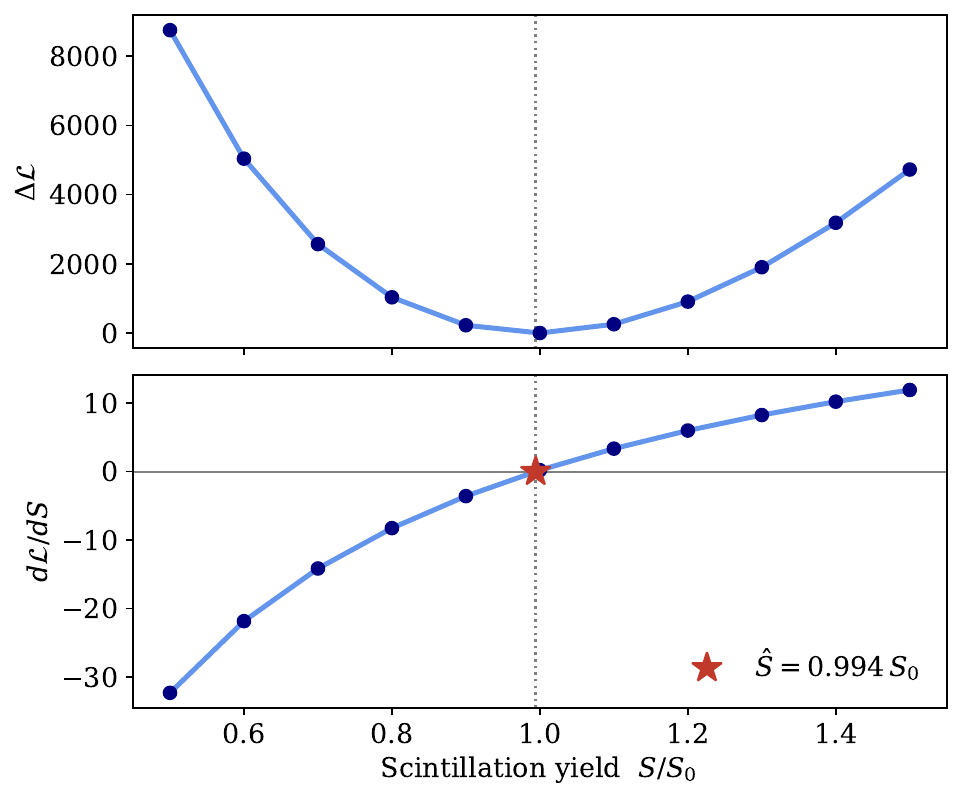}
\end{figure}

Figure~\ref{fig:wbls_example} compares a data-like event with the differentiable prediction produced by \lucid for the fraction of detected light originating from Cherenkov and scintillation processes in each photosensor of a JUNO-sized spherical detector filled with 10\% water based liquid scintillator (WbLS). Using this differentiable prediction, we compute the loss with respect to the nominal scintillation light-yield scale parameter, $S$, for the same data-like event. As shown in Fig.~\ref{fig:scint_diff}, the resulting loss function provides smooth and accurate gradient information through the full Cherenkov plus scintillation \lucid pipeline.

\section{Propagation: algorithms, sampling, and validation}
\label{app:propagation}

Algorithms~\ref{alg:lucid} and~\ref{alg:sampling} are the same photon-transport model run in two modes. Both share the detector geometry, the optical parameters (the scattering and absorption lengths $\lambda_s$ and $\lambda_a$, the surface reflectivity $R$ and specular fraction $f$, and the speed of light $v$ in the medium), and the same free-path, Rayleigh-angle, and specular-versus-diffuse reflection distributions. They differ in exactly one respect: how the two photon-removal processes, detection and absorption, are treated.

The differentiable mode (Algorithm~\ref{alg:lucid}) replaces each removal by its expectation. At every surface it deposits the expected detected charge, the fraction $P_{\mathrm{det}}=P_{\mathrm{reach}}(1-R)$, into the nearby sensors through the soft Gaussian footprint of Sec.~\ref{sec:photon_relaxation}. It then attenuates the surviving ray weight continuously by absorption and reflectivity, so the ray never terminates. This is the implicit-capture variance reduction of Sec.~\ref{sec:photon_prop}. A running DiCE score $\ell$ multiplies every deposit, so the output is differentiable in the physical parameters. The mode converges to the exact expected charge and arrival-time distributions as $N_{\mathrm{rays}}\to\infty$.

The data-like mode (Algorithm~\ref{alg:sampling}) instead samples each removal as a discrete event: the photon reflects, is detected, or is absorbed with the corresponding probabilities. It carries unit weight, strikes a single sensor through a hard hit test, and terminates on detection or absorption. It accumulates no score and is not differentiable. Its output is a set of hits carrying the statistical fluctuations of real detector data, against which the differentiable expectation is validated (Sec.~\ref{sec:data_like}).

\begin{figure*}[t]
\centering
\begin{minipage}{0.98\textwidth}

\algcaption{Differentiable Ray Propagation}
\label{alg:lucid}

\begin{algorithmic}[1]
\Require Ray states $\Psi = \{(\mathbf{x}_i, \mathbf{u}_i, t_i, I_i)\}_{i=1}^N$, parameters $\theta$
\Ensure Deposition list $\mathcal{D}$
\State Initialize $\mathcal{D} \leftarrow \emptyset$, survival $\alpha \leftarrow 1$, score $\ell \leftarrow 0$
\For{$k \leftarrow 1$ \textbf{to} $K$}
\State \textbf{1. Intersection and probabilities}
\State \quad $S_{\text{cand}} \leftarrow \text{GridLookup}(\mathbf{x}_i,\mathbf{u}_i)$;\quad $d_{\text{surf}}, \mathbf{n}, \text{type} \leftarrow \text{AnalyticIntersect}(\mathbf{x}_i,\mathbf{u}_i)$
\State \quad $\mathbf{n} \leftarrow \mathrm{sg}(\mathbf{n})$ \Comment{value kept; drops the normal's curvature gradient, which compounds across bounces}
\State \quad $P_{\text{reach}} \leftarrow e^{-d_{\text{surf}}/\lambda_s}$;\quad $(R,f) \leftarrow \text{Reflect}(\text{type})$;\quad $P_{\text{det}} \leftarrow P_{\text{reach}}(1-R)$

\State \textbf{2. Deposit expected charge} \Comment{implicit capture; ray never terminates}
\State \quad \textbf{for} each candidate sensor $j \in S_{\text{cand}}$:
\State \quad\quad $n_{\text{dep}} \leftarrow I_i\,\alpha_i\,P_{\text{det}}\,e^{-d_{\text{surf}}/\lambda_a}\,w_{\text{geo}}(b_j)$
\State \quad\quad deposit $\;n_{\text{dep}}\,e^{\,\ell - \mathrm{sg}(\ell)}\;$ at $j$, time $t_i + d_{\text{surf}}/v$ \Comment{DiCE factor: $\equiv 1$ forward, score gradient}

\State \textbf{3. Sample the next state}
\State \quad $d_{\text{live}} \leftarrow -\lambda_s\ln(1-\epsilon_d)$;\quad $d \leftarrow \mathrm{sg}(d_{\text{live}})$;\quad $\text{scatter} \leftarrow (d < d_{\text{surf}})$ \Comment{free path; reach if $d \ge d_{\text{surf}}$}
\State \quad $\mathbf{u}_{\text{scat}} \leftarrow \text{Rotate}\!\big(\mathbf{u}_i,\, \text{RayleighInvCDF}(\epsilon_\theta),\, 2\pi\epsilon_\phi\big)$
\State \quad $b \leftarrow [\,\epsilon_b < \mathrm{sg}(f)\,]$;\quad $\mathbf{u}_{\text{refl}} \leftarrow b\,?\,\text{Specular}(\mathbf{u}_i,\mathbf{n}) : \text{Lambertian}(\mathbf{n},\epsilon)$ \Comment{specular w.p.\ $f$, else diffuse}
\State \quad $(\mathbf{x}_i,\mathbf{u}_i,d_{\text{step}}) \leftarrow \text{scatter}\,?\,(\mathbf{x}_i{+}d\,\hat{\mathbf{u}}_i,\, \mathbf{u}_{\text{scat}},\, d) : (\mathbf{x}_i{+}d_{\text{surf}}\hat{\mathbf{u}}_i,\, \mathbf{u}_{\text{refl}},\, d_{\text{surf}})$

\State \textbf{4. Accumulate DiCE score of the realized choices}
\State \quad $\ell \leftarrow \ell + \big(\text{scatter}\,?\,\log\tfrac{1}{\lambda_s}-\tfrac{1}{\lambda_s}\mathrm{sg}(d) : -\tfrac{1}{\lambda_s}\mathrm{sg}(d_{\text{surf}})\big) + b\log f + (1-b)\log(1-f)$

\State \textbf{5. Advance}
\State \quad $t_i \leftarrow t_i + \big(\text{scatter}\,?\,d_{\text{live}} : d_{\text{surf}}\big)/v$ \Comment{arrival time uses the live free path}
\State \quad $\alpha_i \leftarrow \alpha_i\,e^{-d_{\text{step}}/\lambda_a}\,(\text{scatter}\,?\,1 : R)$ \Comment{$(1-R)$ deposited, reflected fraction $R$ continues}
\EndFor
\State \Return $\mathcal{D}$
\end{algorithmic}
\vspace{0.8ex}\hrule
\end{minipage}
\end{figure*}

\begin{figure*}[t]
\centering
\begin{minipage}{0.98\textwidth}

\algcaption{Monte Carlo Ray Propagation (Sampling Mode)}
\label{alg:sampling}

\begin{algorithmic}[1]
\Require Ray states $\Psi = \{(\mathbf{x}_i, \mathbf{u}_i, t_i)\}_{i=1}^N$, parameters $\theta$
\Ensure Hit list $\mathcal{H}$
\State Initialize $\mathcal{H} \leftarrow \emptyset$
\For{$k \leftarrow 1$ \textbf{to} $K$}
\State \textbf{1. Intersection and probabilities}
\State \quad $d_{\text{surf}}, \mathbf{n}, \text{type} \leftarrow \text{AnalyticIntersect}(\mathbf{x}_i,\mathbf{u}_i)$
\State \quad $P_{\text{reach}} \leftarrow e^{-d_{\text{surf}}/\lambda_s}$;\quad $(R,f) \leftarrow \text{Reflect}(\text{type})$
\State \textbf{2. Sample the outcome and advance}
\If{$\epsilon_1 \ge P_{\text{reach}}$} \Comment{scatter in the volume}
\State \quad $d \leftarrow -\lambda_s\ln\!\big(1 - \epsilon_d(1 - e^{-d_{\text{surf}}/\lambda_s})\big)$ \Comment{scatter distance (truncated exp)}
\State \quad $\mathbf{x}_i \leftarrow \mathbf{x}_i + d\,\hat{\mathbf{u}}_i$;\quad $\mathbf{u}_i \leftarrow \text{Rotate}\!\big(\mathbf{u}_i, \text{RayleighInvCDF}(\epsilon_\theta), 2\pi\epsilon_\phi\big)$;\quad $t_i \leftarrow t_i + d/v$
\State \quad \textbf{if} $\epsilon_a \ge e^{-d/\lambda_a}$ \textbf{then return} \Comment{absorbed en route}
\Else \Comment{reach the surface}
\State \quad $t_i \leftarrow t_i + d_{\text{surf}}/v$
\State \quad \textbf{if} $\epsilon_a \ge e^{-d_{\text{surf}}/\lambda_a}$ \textbf{then return} \Comment{absorbed en route}
\If{$\epsilon_R < R$} \Comment{reflect: specular w.p.\ $f$, else diffuse}
\State \quad $\mathbf{x}_i \leftarrow \mathbf{x}_i + d_{\text{surf}}\hat{\mathbf{u}}_i + \delta\hat{\mathbf{n}}$;\quad $\mathbf{u}_i \leftarrow [\epsilon_b < f]\,?\,\text{Specular}(\mathbf{u}_i,\mathbf{n}) : \text{Lambertian}(\mathbf{n},\epsilon)$
\ElsIf{$\text{type} = \text{Sensor}$} \Comment{detected}
\State \quad $\mathcal{H}.\text{append}(\{\text{sensor}: \text{id}(\mathbf{n}),\ \text{time}: t_i\})$;\quad \textbf{return}
\Else \Comment{wall, not reflected: absorbed}
\State \quad \textbf{return}
\EndIf
\EndIf
\EndFor
\State \Return $\mathcal{H}$
\end{algorithmic}
\vspace{0.8ex}\hrule
\end{minipage}
\end{figure*}

\subsection{Reparameterized Functions}
\label{app:reparam}

Here we detail the differentiable mappings $f(\epsilon; \theta)$ used to transform uniform noise $\epsilon \sim \mathcal{U}[0, 1]$ into physical quantities, as implemented in our JAX-based simulation.

\subsubsection{Volumetric Scattering Distance}
To sample a scattering distance $d_s$ strictly within the detector volume bounded by $d_{\mathrm{surf}}$, we employ a truncated exponential distribution. The normalized CDF is inverted to yield the differentiable mapping:
\begin{equation}
    d_s(\epsilon, \lambda_s) = -\lambda_s \ln\left( 1 - \epsilon \left( 1 - e^{-d_{\mathrm{surf}}/\lambda_s} \right) \right).
\end{equation}
This formulation ensures $0 \le d_s \le d_{\mathrm{surf}}$ while preserving gradient flow from the interaction point back to the scattering length $\lambda_s$.

\subsubsection{Rayleigh Scattering Angle}
For Rayleigh scattering the phase function is $p(\mu) \propto 1 + \mu^2$ in the cosine of the scattering angle $\mu = \cos\theta \in [-1, 1]$. Normalizing over $[-1, 1]$ gives $p(\mu) = \tfrac{3}{8}\,(1 + \mu^2)$, whose cumulative distribution is
\begin{equation}
F(\mu) = \int_{-1}^{\mu} p(\mu')\,d\mu'
       = \tfrac{1}{8}\left( \mu^3 + 3\mu + 4 \right).
\end{equation}
Applying inverse-transform sampling, we set $F(\mu) = \epsilon$ with $\epsilon \sim \mathcal{U}[0, 1]$, which rearranges to the depressed cubic
\begin{equation}
    \mu^3 + 3\mu + 4(1 - 2\epsilon) = 0.
\end{equation}
The map $F$ is strictly monotonic and this cubic has a single real root for every $\epsilon$; that root is the sampled $\mu$. Standard rejection sampling of $p(\mu)$ would be non-differentiable, so instead we solve the cubic analytically with Cardano's formula, obtaining a closed-form, differentiable mapping $\mu(\epsilon)$. The scattered direction is completed with an azimuth $\phi = 2\pi\epsilon'$, $\epsilon' \sim \mathcal{U}[0, 1]$, and rotated into the lab frame about the incident ray direction. The Rayleigh phase function itself carries no tunable parameter; nonetheless, because $\mu(\epsilon)$ and $\phi(\epsilon')$ are closed-form and the noise $(\epsilon, \epsilon')$ is held fixed, this reparameterization keeps the scattered direction differentiable and lets gradients propagate through it into the local scattering frame.

\subsubsection{Lambertian Reflection}
We simulate diffuse reflection by mapping two noise samples $\epsilon_1, \epsilon_2$ to a cosine-weighted hemisphere. The local vector is constructed as:
\begin{equation}
    \mathbf{u}_{\mathrm{local}} = \left[ \sqrt{\epsilon_1}\cos(2\pi\epsilon_2), \, \sqrt{\epsilon_1}\sin(2\pi\epsilon_2), \, \sqrt{1-\epsilon_1} \right]^T.
\end{equation}
This vector is then rotated into the global frame using a basis derived from the surface normal $\mathbf{n}$, allowing gradients to propagate through the surface orientation.

\subsection{Two-Channel Scattering and the Mie Phase Function}
\label{app:mie}

The scattering model can include a strongly forward-peaked Mie channel alongside Rayleigh scattering. The two channels combine by rate. With a Rayleigh length $\lambda_R$ (the scattering length $\lambda_s$ of Sec.~\ref{sec:photon_prop}) and a Mie length $\lambda_M$, the total scattering rate is $\mu_{\rm tot} = 1/\lambda_R + 1/\lambda_M$. A given scatter is of Mie type with probability $p_{\rm Mie} = (1/\lambda_M)/\mu_{\rm tot}$. The free path of Sec.~\ref{sec:photon_prop} is then drawn from an exponential of rate $\mu_{\rm tot}$, and the Rayleigh-versus-Mie assignment is a Bernoulli draw with probability $p_{\rm Mie}$.

The Mie deflection angle is drawn from a Henyey--Greenstein phase function of asymmetry parameter $g$, which for forward-peaked Mie scattering we take in the range $g \in [0, 1)$; written in terms of the scattering-angle cosine $\cos\theta$,
\begin{equation}
\label{eq:hg}
p_{\rm HG}(\cos\theta) = \frac{1}{2}\,\frac{1 - g^{2}}{\left(1 + g^{2} - 2 g \cos\theta\right)^{3/2}}.
\end{equation}
Inverse-transform sampling of Eq.~\eqref{eq:hg} gives a closed-form, differentiable map from a uniform noise sample $u \sim \mathcal{U}[0, 1]$,
\begin{equation}
\label{eq:hg_sample}
\cos\theta(u) = \frac{1}{2g}\left[\,1 + g^{2} - \left(\frac{1 - g^{2}}{\,1 - g + 2 g u\,}\right)^{2}\,\right],
\end{equation}
which reduces to isotropic scattering as $g \to 0$. This is the direct analogue, for the Mie channel, of the Cardano solution used for the Rayleigh angle (Appendix~\ref{app:reparam}); the azimuth is drawn uniformly and the direction rotated into the laboratory frame as before. Although this map would support reparameterization, here the sampled cosine is held fixed with respect to $g$ by a stop-gradient. The gradient of $g$ is carried entirely by the angle score below, so no double counting arises.

Because $\lambda_R$, $\lambda_M$, and $g$ now enter the sampling probabilities, their gradients are carried by the free-path and angle scores of Appendix~\ref{app:propagation}. With $\mathrm{sg}(\cdot)$ the stop-gradient operator and $d$ the sampled free path, the free-path score on a scatter is $\ell_{\rm path} = \log\mu_{\rm tot} - \mu_{\rm tot}\,\mathrm{sg}(d)$, which carries $\lambda_R$ and $\lambda_M$ through $\mu_{\rm tot}$. The angle score gains the mixture term
\begin{equation}
\ell_{\rm ang} =
\begin{cases}
\log p_{\rm Mie} + \log p_{\rm HG}\!\big(\mathrm{sg}(\cos\theta); g\big), & \text{Mie},\\[2pt]
\log(1 - p_{\rm Mie}) + \log p_{\rm Ray}\!\big(\mathrm{sg}(\cos\theta)\big), & \text{Rayleigh},
\end{cases}
\end{equation}
where $\log p_{\rm HG}(\cos\theta;g) = \log(1-g^{2}) - \log 2 - \tfrac{3}{2}\log(1 + g^{2} - 2 g \cos\theta)$ from Eq.~\eqref{eq:hg} and $p_{\rm Ray}(\cos\theta) \propto (1 + \cos^{2}\theta)$ carries no parameter. The DiCE surrogate then makes $\lambda_M$ and $g$ differentiable exactly as for the single-channel case. Both scattering lengths additionally enter pathwise through the deposit factor $\exp(-\mu_{\rm tot}\,d_{\rm surf})$, in which $\mu_{\rm tot}$ depends on $\lambda_R$ and $\lambda_M$ alike.

We do not calibrate $\lambda_M$ or $g$ in the water results of Sec.~\ref{sec:calibration}. For the Super-Kamiokande water optical model the forward-Mie length is of order $10^{4}$~m, so the Mie channel contributes at the sub-percent level to the charge pattern. We hold $\lambda_M$ and $g$ fixed at their nominal values and leave their calibration for future work. The same two-channel treatment carries over unchanged to strongly scattering media, where the Mie channel instead dominates.

\subsection{Spatial Overlap Integral}
\label{app:overlap}
The photon relaxation weight $w_{\rm geo}$ represents the probability of a Gaussian photon packet striking a circular sensor of radius $r$. We define the integral in polar coordinates $(\rho, \phi)$ centered on the sensor, where the photon is incident at a distance $b$ from the center:
\begin{equation}
\mathcal{I}(b) = \int_{0}^{r} \int_{0}^{2\pi} \frac{\rho}{2\pi\sigma^2} \exp\left( -\frac{\rho^2 + b^2 - 2\rho b \cos\phi}{2\sigma^2} \right) d\phi \, d\rho.
\end{equation}
Evaluating this double integral during propagation is computationally expensive. Instead, we precompute the function $\mathcal{I}(b)$ and store it as a 1D lookup table. To maintain high precision while minimizing memory usage, we do not use a uniform grid. The function $\mathcal{I}(b)$ changes most rapidly near the sensor edge ($b \approx r$), where it transitions from $\approx 1$ to $\approx 0$. We therefore construct a non-uniform grid for $b$ with three distinct regions:

\begin{enumerate}
\item Inner Sparse Region ($0 \le b < r - 3\sigma$): The overlap is near-unity and flat; we use sparse sampling here.
\item Transition Region ($|b - r| \le 3\sigma$): The overlap drops sharply; we use a denser grid to capture the gradient accurately.
\item Outer Sparse Region ($b > r + 3\sigma$): The overlap tails off to zero; we use sparse sampling up to a cutoff distance (e.g., $10r$).
\end{enumerate}

We typically use 150 points for the transition region and 50 points for the inner and outer sparse regions accordingly. During simulation, the value of $w_{\rm geo}$ is retrieved via linear interpolation of this lookup table. This operation is fully differentiable, allowing gradients to propagate through the lookup values with respect to the impact parameter $b$.

\subsection{Gradient Flow in the Propagation Loop}
\label{app:grad_truncation}

Reverse-mode automatic differentiation propagates exact gradients through the full propagation loop, including the geometric trajectories of scattered and reflected rays, so the ray position and direction gradients flow through all $K$ iterations. The only term we drop is the derivative of the surface normal at reflection. On a curved surface the normal depends on the hit point, and differentiating the reflected direction through that dependence introduces a term proportional to the local surface curvature, which compounds multiplicatively across successive bounces and destabilizes the gradient at large propagation depth. We therefore hold the normal fixed under differentiation. Its value still enters the reflected direction and the reflectance, both of which remain differentiable through the incident ray, and the deposited charge and DiCE score are unaffected. No depth-based truncation of the loop is then required.

\subsection{Convergence of Propagation Depth}
\label{app:Kconvergence}

The computational cost of the simulation scales linearly with the maximum number of propagation iterations, $K$. It is therefore crucial to select a value of $K$ that minimizes computational overhead while ensuring that the total neglected photon intensity remains negligible.

To determine the optimal cutoff, we simulated the propagation of light from an isotropic point source placed at the center of two detector geometries: an SK-like variant geometry ($20\,\text{m} \times 40\,\text{m}$) and an HK-like variant geometry ($35\,\text{m} \times 70\,\text{m}$). We utilized a representative set of optical parameters: a scattering length $\lambda_s = 40\,\text{m}$, an absorption length $\lambda_a = 200\,\text{m}$, and a surface reflection rate $R = 0.25$.

\begin{figure}[htbp]
\centering
\caption{The fraction of initial photon intensity remaining active in the simulation as a function of the number of propagation iterations ($K$).}
\label{fig:intensity_decay}
\includegraphics[width=0.48\textwidth]{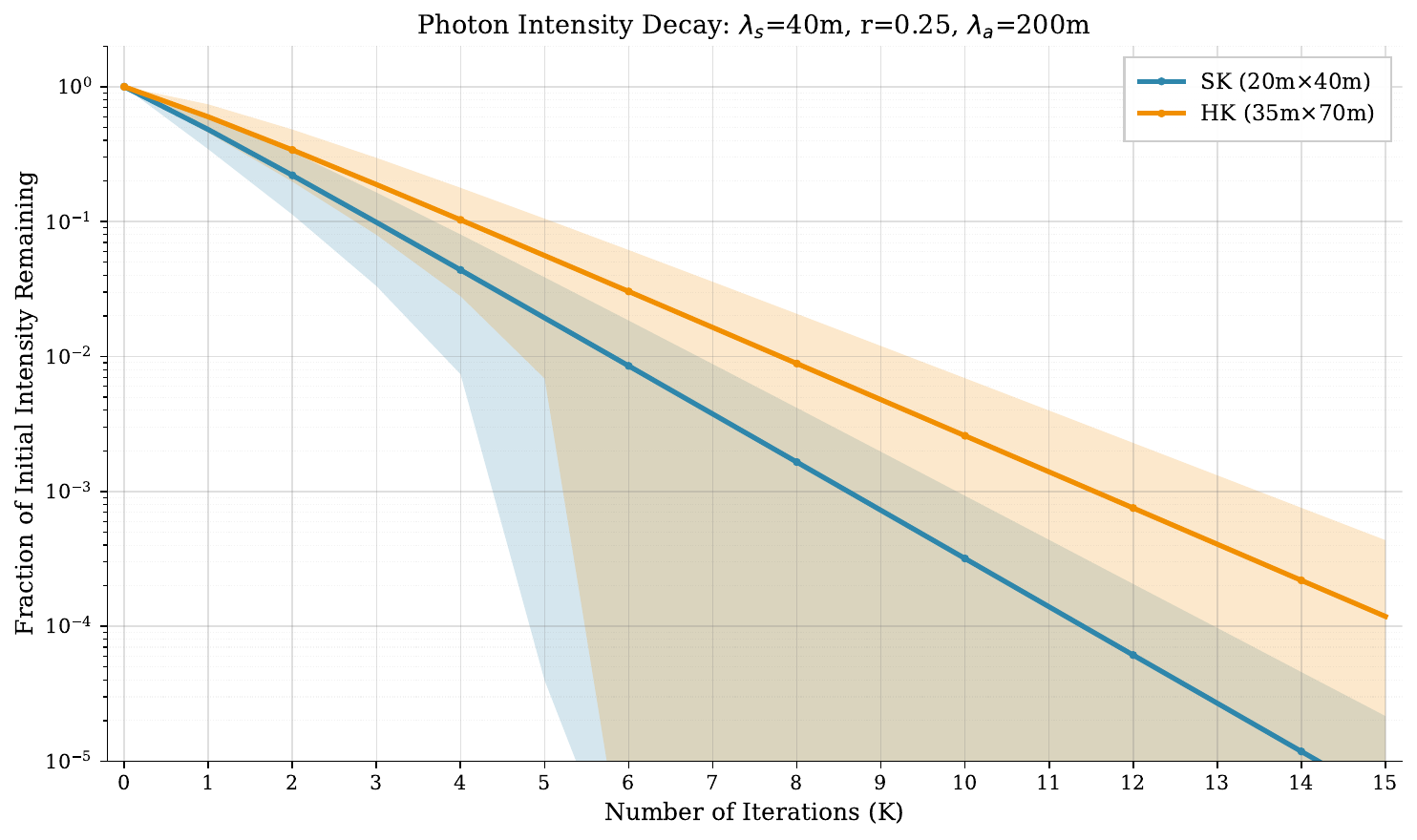}
\end{figure}

Figure~\ref{fig:intensity_decay} shows the fraction of the initial photon intensity remaining in the simulation queue as a function of the iteration index $k$. The intensity decays exponentially as photons are absorbed by the medium or detected by the photosensors. For the SK-like geometry, we observe that at $K=8$, the average remaining intensity drops below $0.2\%$. This threshold serves as our convergence criterion. Larger detector volumes, such as the HK-like geometry, exhibit a slightly slower decay rate due to the longer average path lengths between surface interactions, yet they follow a similar convergence profile.

\section{The Gauss--Newton Fit}
\label{app:optimizer}

\subsection{The Gauss--Newton matrix and the choice of loss}

Calibration and reconstruction are both parameter-estimation problems, in which the parameters are varied until the predicted per-sensor charges $\mu_j(\bm\theta)$ reproduce the observed charges $q_j$. The estimate is the minimizer of a loss function that quantifies the per-sensor discrepancy between prediction and observation. The Gauss--Newton optimizer of Sec.~\ref{sec:gauss_newton} requires the gradient and curvature of this loss. Counting data admit several standard loss functions. We derive the curvature for the Poisson likelihood, and then show that the common alternatives share it, so that a single optimizer applies to all of them.

The canonical loss for photon counts is the Poisson negative log-likelihood. For an observed count $q_j$ and predicted mean $\mu_j(\bm\theta)$ (Eq.~\ref{eq:hit_charge}), it reads, up to a $\bm\theta$-independent constant,
\begin{equation}
  \mathcal{L} = \sum_j\big[\,\mu_j - q_j\log\mu_j\,\big],
  \label{eq:poisson_nll}
\end{equation}
with gradient and Hessian
\begin{equation}
  \frac{\partial\mathcal{L}}{\partial\theta_i}
    = \sum_j \frac{\mu_j-q_j}{\mu_j}\,\frac{\partial\mu_j}{\partial\theta_i},
  \label{eq:gn_gradient}
\end{equation}
\begin{equation}
  \frac{\partial^2\mathcal{L}}{\partial\theta_i\partial\theta_k}
    = \sum_j \frac{q_j}{\mu_j^{2}}\,
        \frac{\partial\mu_j}{\partial\theta_i}\frac{\partial\mu_j}{\partial\theta_k}
    + \sum_j \frac{\mu_j-q_j}{\mu_j}\,
        \frac{\partial^2\mu_j}{\partial\theta_i\partial\theta_k}.
  \label{eq:gn_hessian}
\end{equation}
The Hessian comprises two contributions. The first depends only on the first derivatives of the prediction. The second is proportional to the residual $(\mu_j-q_j)$ and involves the second derivatives $\partial^2\mu_j/\partial\theta_i\partial\theta_k$. These are costly to evaluate, and need not be positive-definite, so that the exact Hessian can furnish an ascent direction. Substituting the model means for the observed counts, $q_j\!\to\!\mu_j$, eliminates the second contribution identically, because the residual then vanishes. The resulting curvature is independent of the particular realization of the data. This \emph{expected} curvature is the Gauss--Newton matrix of Eq.~\ref{eq:fisher}.

This matrix is the \emph{Fisher information} of the model, which quantifies the sensitivity of the data to each parameter and to each combination of parameters. With the score $\partial\log P/\partial\theta_i = \sum_j(q_j/\mu_j-1)\,\partial\mu_j/\partial\theta_i$ and the Poisson variance $\mathrm{Var}(q_j)=\mu_j$,
\begin{equation}
  \mathcal{I}_{ik}
    = \mathbb{E}\!\left[\frac{\partial\log P}{\partial\theta_i}
                        \frac{\partial\log P}{\partial\theta_k}\right]
    = \sum_j \frac{\mathrm{Var}(q_j)}{\mu_j^{2}}\,
        \frac{\partial\mu_j}{\partial\theta_i}\frac{\partial\mu_j}{\partial\theta_k}
    = F_{ik}.
  \label{eq:fisher_identity}
\end{equation}
It is positive-semidefinite by construction, so the Gauss--Newton step is never an ascent direction. Its inverse is the Cram\'er--Rao bound, treated as indicative in Sec.~\ref{sec:gauss_newton}.

The Poisson likelihood is one of several loss functions that share this Gauss--Newton matrix. The identity $\partial\sqrt{\mu}/\partial\theta = (\partial\mu/\partial\theta)/2\sqrt{\mu}$ relates $F$ to a least-squares problem in the transformed variable $\sqrt{\mu}$. For Poisson data $\sqrt{q}$ is the variance-stabilizing transform, its variance approaching $\tfrac14$ independently of $\mu$. A least-squares fit in $\sqrt{q}$ therefore assigns each sensor an approximately equal weight. The objective $2\sum_j(\sqrt{q_j}-\sqrt{\mu_j})^{2}$ has Gauss--Newton matrix equal to $F$. A further standard choice is the Neyman $\chi^2$, $\sum_j(\mu_j-q_j)^2/q_j$, the least-squares statistic in which each term is weighted by its observed count. It is linear in the predicted charge. Its Gauss--Newton matrix $2\sum_j(\partial\mu_j/\partial\theta_i)(\partial\mu_j/\partial\theta_k)/q_j$ reduces to $2F$ near the minimum, where $q_j\approx\mu_j$. The three matrices are each proportional to $F$, and the constant cancels in the Newton direction $F^{-1}\bm g$. Near a shared minimum they therefore take the same second-order step. Away from it they weight the residual differently, so on finite, noisy data their minima, and their small biases, need not coincide. The loss function is thus a modeling choice matched to the problem, while the second-order machinery is unchanged.

\subsection{The per-sensor efficiencies}

Partition the parameters into the bulk optical properties $\bm\phi$, few in number, and the per-sensor efficiencies $\{k_s\}$, one per sensor and therefore many. Each $k_s$ is a dimensionless factor multiplying the overall quantum efficiency at its own sensor alone, so across the sources $c$ that illuminate it the prediction is $\mu_{sc} = k_s\,m_{sc}(\bm\phi)$, with $m_{sc}$ the charge at unit efficiency. Requiring the predicted charge at each sensor to match the charge observed there, summed over the sources, determines the efficiency at given $\bm\phi$,
\begin{equation}
  k_s(\bm\phi) = \frac{\sum_c q_{sc}}{\sum_c m_{sc}(\bm\phi)}.
  \label{eq:khat}
\end{equation}
This is the maximum-likelihood value for Poisson counts. It is linear in the data and therefore unbiased, which is why we adopt it for every choice of loss.

Substituting Eq.~\ref{eq:khat} into the prediction leaves a model in $\bm\phi$ alone. The Gauss--Newton matrix of Eq.~\ref{eq:fisher} is assembled over those parameters only, from the derivatives of $k_s m_{sc}$ taken at the current efficiencies, so the linear system solved at each iteration has their dimension rather than the number of sensors. Evaluating Eq.~\ref{eq:khat} costs $\mathcal{O}(N_{\mathrm{sensors}})$ per iteration, and at convergence it supplies the recovered efficiencies, which are results of the fit rather than quantities discarded from it.

Two approximations enter. Eq.~\ref{eq:khat} is the Poisson condition, while the step in $\bm\phi$ minimizes the Neyman $\chi^2$; the two coincide where the model reproduces the data, and differ at the scale of the statistical fluctuations elsewhere. The response of Eq.~\ref{eq:khat} to $\bm\phi$ is also not propagated through the derivative. Carrying it would add a second term, $\partial(k_s m_{sc})/\partial\bm\phi = k_s\big[\partial m_{sc}/\partial\bm\phi-(m_{sc}/M_s)\,\partial M_s/\partial\bm\phi\big]$ with $M_s=\sum_c m_{sc}$, which subtracts from each sensor the part of the response common to all of its sources. That common part is what the efficiency absorbs, so only the redistribution of light among the sources carries information about $\bm\phi$. With limited photon statistics the efficiencies can absorb part of what the optical parameters would otherwise account for, biasing those that produce only small changes in the collected light. The effect diminishes as the photon budget grows.

One direction must be fixed before the system can be inverted. A global rescaling $k_s\!\to\!c\,k_s$ leaves every $\mu_s$ unchanged when a compensating change of the overall normalization absorbs it, and that normalization is carried by the overall quantum efficiency and the source brightness. The problem therefore has an exact flat direction. The gauge condition $\langle\log k_s\rangle = 0$ removes it, assigning the normalization to the overall quantum efficiency rather than to the per-sensor factors.

\subsection{The calibration fit}

Nineteen parameters are fit. Fifteen are the medium properties, a scattering length, an absorption length, and an overall quantum efficiency at each of the five wavelengths, left independent so that no spectral model is imposed on the recovered curve. The remaining four are the wall and sensor reflectivities together with their specular fractions, shared across all wavelengths because the reflection model carries no wavelength dependence. The Mie length and asymmetry parameter are held fixed, so the forward Mie channel is inactive in these fits and the recovered scattering length is the Rayleigh length alone.

The parameters are fit in transformed coordinates. The fifteen medium properties are fit in log space, which enforces positivity and puts their widely different magnitudes on a common footing. The reflectivities are fit as the logarithm of the total reflectivity and the logit of the specular fraction, which keeps the fraction in the unit interval without explicit bounds and gives the total reflectivity a direct gradient rather than requiring its two components to move together.

The detector is illuminated by eight sources: a laser at the top end cap directed along the axis, and seven isotropic sources at the center of the volume and at $\pm 10$~m along each coordinate axis. Each is fired at all five wavelengths, giving forty source and wavelength configurations, each simulated with $10^{6}$ rays and $K=12$ propagation steps. The reference data are the expected response at the injected parameters, averaged over eight ray realizations, and are identical for every starting point, so that the spread across starts measures the variability of the optimizer rather than of the data. The injected per-sensor efficiencies carry a $5\%$ lognormal spread.

The residual is the Neyman $\chi^2$ of Eq.~\ref{eq:chi2_loss}, with the weight floored at one percent of the mean observed charge so that the dimmest sensors cannot dominate. The Jacobian is obtained by forward-mode differentiation of that residual, averaged over eight independent ray realizations and rebuilt every twentieth iteration, while the residual itself is refreshed at every iteration. Averaging matters more than freshness here: the specular fractions enter through the score-function estimator of Sec.~\ref{sec:photon_prop} rather than pathwise, and the noise in their Jacobian is comparable to their curvature, so the same noisy realization entering both factors of $J^{\mathsf T}J$ inflates the curvature and shortens the step. Averaging suppresses that; rebuilding more often does not.

The fit runs for $600$ iterations from three starting points, each perturbed uniformly from the injected values by up to $0.4$ in the fit coordinates, widened to $2.0$ for the scattering lengths, a factor of about seven, and $1.2$ for the reflection parameters. Each step is clipped at $0.5$ in log units. The reported answer is the mean of the last $50$ iterates.

\subsection{The track fit}

Nine parameters are carried for the seven degrees of freedom of Sec.~\ref{sec:tracking}. Rather than a transformation, they are preconditioned by fixed characteristic scales, $50$~MeV in energy, $0.2$~m in position, $0.2$~ns in time, and $0.02$ in each direction component, applied before the linear solve. This is not cosmetic. The undamped metric is dominated by the position block, which exceeds the energy block by about five orders of magnitude, and without the rescaling the energy direction does not move at all. The scales correspond to a uniform relative step of about two percent, which is what a logarithmic parameterization would supply were the position, time, and direction components not signed.

Both observables enter. The charge term is the Poisson negative log-likelihood of Eq.~\ref{eq:counts_loss}, left unnormalized, since the total predicted charge is how the energy enters the prediction and normalizing it removes the energy constraint entirely. The timing term is the first-arrival order statistic of Eq.~\ref{eq:time_loss}, with its kernel width matched to the transit-time spread used to generate the data. The metric is the sum of the two corresponding blocks, the charge block weighted by the inverse predicted mean and the timing block the covariance of its score, both assembled by forward-mode differentiation over eight ray realizations and rebuilt every eighth iteration. The gradient, unlike the calibration fit, is taken in reverse mode from the scalar loss at every iteration. The metric therefore acts purely as a preconditioner, and the descent direction is exact whatever the state of the metric.

The energy is treated specially. It enters the charge prediction both through the total light yield and through the shape of the predicted pattern, and because the emission surrogate reproduces that shape only approximately, differentiating through the shape lets a small misspecification bias the energy by way of its degeneracy with the vertex along the track. We therefore route the energy gradient through the total predicted charge alone, holding the predicted shape fixed. This uses the same information a total light-yield estimate would, removes the bias, and slightly narrows the spread. It also steepens the energy gradient enough to require a trust region, and the step is accordingly clipped at three in the scaled coordinates.

The step is amplified by a factor annealed linearly from $4.0$ to $1.5$ over $150$ iterations. This factor is not a learning rate in the usual sense: a consistent Gauss--Newton step systematically undershoots, so values above one are expected. Large early steps converge quickly while small late ones cannot push a converged fit back out of its basin. Any step producing a non-finite parameter or gradient is discarded and the previous iterate retained. The reported answer is the mean of the last $40$ iterates.

The fit is started from the seeds of Appendix~\ref{app:track_initial_guesses}. Three are built, one from the charge grid, one from multilateration of the arrival times, and one fusing the transverse component of the second with the longitudinal component of the first. One is selected before the fit by data loss, with the charge-grid seed preferred unless another improves on it by more than one percent. A plain minimum over the three would over-select the time-based seed, because the loss minimum does not sit exactly at truth and its forward-biased basin can score marginally lower while giving a worse vertex.

\subsection{Fitting against a Monte Carlo forward model}

The predicted charge in Eq.~\ref{eq:calib_loss} is estimated from a finite set of simulated rays, so it carries a Monte Carlo error that changes whenever those rays are redrawn. Writing the summand of that objective at a realization $\xi$ of the rays as
\begin{equation}
\label{eq:calib_resid}
r_{sc}(\bm\phi;\xi) = \frac{k_s(\bm\phi;\xi)\,m_{sc}(\bm\phi;\xi) - q_{sc}}{\sqrt{q_{sc}}},
\end{equation}
the two pieces of the step of Sec.~\ref{sec:calibration_method} are
\begin{equation}
\label{eq:calib_step}
\bar{J} = \frac{1}{B}\sum_{b=1}^{B}
  \frac{\partial \bm r(\bm\phi;\xi'_b)}{\partial\bm\phi},
\qquad
\bm g = \bar{J}^{\mathsf T}\,\bm r(\bm\phi;\xi),
\qquad
F = \bar{J}^{\mathsf T}\bar{J},
\end{equation}
with $B = 8$ and the realizations $\{\xi'_b\}$ drawn independently of $\xi$. Three features of this arrangement are deliberate, and each departs from what one would do with a deterministic model.

The realization $\xi$ is redrawn at every iteration rather than held fixed. Freezing it would make the objective a fixed function of $\bm\phi$ and give a smoother trajectory, but the fit would then converge to the minimum of one noise realization rather than of its expectation, and that displacement does not average away with further iterations. Holding the stream fixed was measured to leave the recovered parameters roughly five times further from their injected values at the same cost.

The realizations $\{\xi'_b\}$ entering $\bar{J}$ are drawn independently of $\xi$. Because the gradient is a product of the two, its expectation factorizes only when they are independent, and a shared realization contributes a covariance term that enters the step as a bias rather than as noise. Drawing the rays separately was measured to suppress that term at no cost in variance. It is also why the gradient is not obtained by differentiating a scalar loss, since automatic differentiation of a single evaluation shares the draw by construction.

The reported estimate is the tail average
\begin{equation}
\label{eq:polyak}
\hat{\bm\phi} = \frac{1}{W}\sum_{t=T-W+1}^{T}\bm\phi_t ,
\end{equation}
with $T = 600$ and $W = 50$ for the calibration and $T = 150$ and $W = 40$ for the track fit. The iterate does not converge to a point but wanders on a floor set by the Monte Carlo error, so a single iterate is a draw from a distribution rather than an estimate, and averaging the tail is what makes the result reproducible.

The track fit shares the first and third of these. It does not face the second, because its gradient is taken in reverse mode from the scalar loss at every iteration while the metric is assembled separately, so the two are never built from the same object.

\subsection{The damped step}

Both fits take the same step. The Gauss--Newton matrix is damped as $F + \gamma\,\mathrm{diag}(F) + \mu\,\mathrm{median}(\mathrm{diag}\,F)\,\mathbb{I}$ with $\gamma = 0.01$ and $\mu = 0.1$, and the resulting system is solved directly. The first term is per-parameter and scale free, damping stiff and soft directions by the same relative amount. The second is a single absolute floor, which bounds the directions the data constrain least and keeps the solve well posed.

\section{Track Initial Guesses}
\label{app:track_initial_guesses}

\textbf{Stage 0: Initial Energy Estimation.} An initial estimate of the particle energy is obtained by minimizing the difference between the observed and predicted total photoelectron counts. The energy scan evaluates 57 trial values spanning $[200, 3000]$~MeV in $50$~MeV steps. The same window is used for every sample, so the seeding is not tuned to the energy under study. During this stage, the track origin is fixed at the detector center, and the track direction is set to the arbitrary unit vector $\mathbf{d} = (1/\sqrt{3}, 1/\sqrt{3}, 1/\sqrt{3})$. Since the evaluation metric depends only on the total signal normalization and not on the detailed charge pattern, the specific choice of position and direction is not critical. The chosen configuration, however, ensures that counts are distributed across multiple detector surfaces for most geometries. The resulting energy is shared by both vertex seeds below.\\

\textbf{Stage 1: Position and Time Offset Search.} A hierarchical grid search determines the vertex position and time offset $t_0$ simultaneously. The algorithm evaluates 9 uniformly spaced $t_0$ values in the range $[-20, 20]$~ns. For each $t_0$, a complete three-dimensional hierarchical position search is performed as follows:
\begin{itemize}\setlength\itemsep{-0.3em}
\item The search begins at the detector center with a cubic grid spanning the detector volume.
\item Each spatial dimension is divided into $n_{\text{div}} = 5$ intervals with an initial grid spacing $L$.
\item The time loss is evaluated at every grid point.
\item The search is refined hierarchically over 6 levels.
\item At each level, the grid is re-centered on the best position from the previous iteration, with the search region half-size set to $L/2$, and a new spacing defined as $L_{\text{new}} = L/n_{\text{div}}$.
\end{itemize}
The position and $t_0$ combination yielding the minimum loss over all tested $t_0$ values is selected as the best estimate. This is the charge-grid vertex.\\

\textbf{Stage 2: Direction Search.} A hierarchical cone-based search determines the particle direction through the following steps:
\begin{itemize}\setlength\itemsep{-0.3em}
    \item Global sampling is performed with 8 divisions in polar angle $\theta \in [0, \pi]$ and 16 divisions in azimuthal angle $\phi \in [0, 2\pi)$, forming a uniform angular grid.
    \item For each sampled direction, the loss $\mathcal{L}_{\mathrm{counts}}$ between observed and simulated photon counts is evaluated.
    \item Around the best-performing direction from the global scan, a cone with opening angle $\alpha_{\text{max}} = 90^\circ$ is generated. It is populated by four concentric rings of points, with the number of points per ring increasing with radius to maintain uniform angular spacing.
    \item The search proceeds over 3 levels, the cone opening angle being halved at each level.
\end{itemize}
The direction minimizing $\mathcal{L}_{\mathrm{counts}}$ across all tested directions is selected as the best estimate.\\

\textbf{Stage 3: Complementary Vertex and Seed Fusion.} The charge pattern fixes the direction well but is degenerate in where along the track the vertex sits. A second, independent vertex is therefore obtained by multilateration of the first-arrival times, which breaks that degeneracy by triangulation and requires no forward simulation. Outliers from scattered and reflected light are suppressed by restricting the fit to the brightest sensors within a time gate. Stage 2 is then repeated for this vertex, yielding a second complete seed. A third candidate is formed by fusing the two, taking the transverse vertex component from the time-based seed and the longitudinal component from the charge-grid seed, the direction and energy from the charge-grid seed, and $t_0$ as the mean of the two, whose biases are of opposite sign. The fit is started from the candidate with the lowest data loss, preferring the charge-grid seed unless another improves on it by more than 1\%.

\vfill
\newpage

\end{document}